\documentclass[12pt]{article} 
\usepackage{graphicx}
\usepackage{epsfig} 
\usepackage{latexsym} 
\usepackage{latexsym}
\usepackage{amssymb} 
\usepackage{amsmath} 
\usepackage{cite}
\newcommand{\labell}[1]{\label{#1}}
\newcommand{\reef}[1]{(\ref{#1})}

\def\Tr{{\rm Tr}}

\DeclareSymbolFont{AMSb}{U}{msb}{m}{n}
\DeclareMathSymbol{\IN}{\mathbin}{AMSb}{"4E}
\DeclareMathSymbol{\IZ}{\mathbin}{AMSb}{"5A}
\DeclareMathSymbol{\IR}{\mathbin}{AMSb}{"52}
\DeclareMathSymbol{\Q}{\mathbin}{AMSb}{"51}
\DeclareMathSymbol{\II}{\mathbin}{AMSb}{"49}
\DeclareMathSymbol{\IC}{\mathbin}{AMSb}{"43}
\DeclareMathSymbol{\IP}{\mathbin}{AMSb}{"50}
\DeclareMathSymbol{\IH}{\mathbin}{AMSb}{"48}
\DeclareMathSymbol\IA{\mathalpha}{AMSb}{"41}
\DeclareMathSymbol\IS{\mathalpha}{AMSb}{"53}

\def\Q{{\cal Q}}

\begin{document}

\begin{flushright}
USC-04-05
\end{flushright}
{\flushleft\vskip-1.35cm\vbox{\psfig{figure=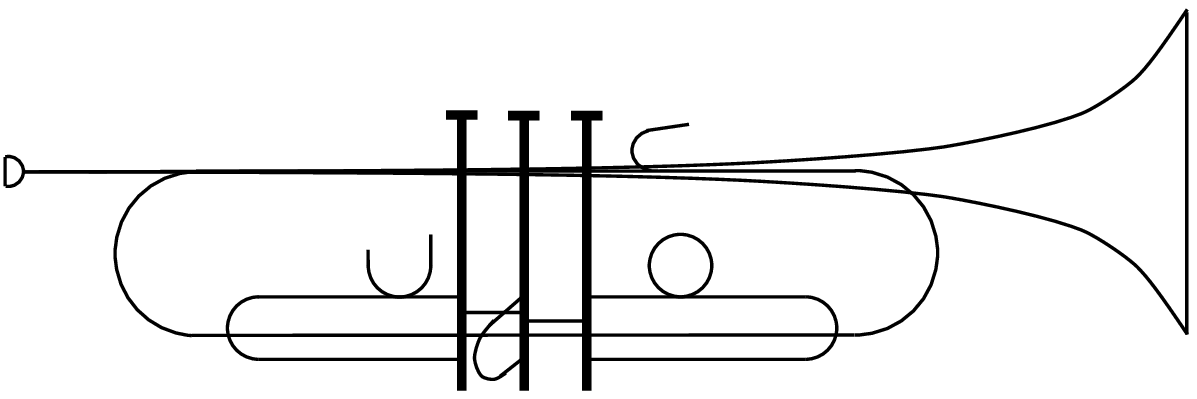,height=0.45in}}}
\bigskip
\begin{center} {\Large \bf Tachyon Condensation,}



\bigskip

{\Large\bf  Open--Closed Duality,}

\bigskip

{\Large\bf Resolvents,}

\bigskip

{\Large\bf and}
 
\bigskip

{\Large\bf Minimal Bosonic and Type 0 Strings}

\end{center}

\bigskip \bigskip \bigskip

\centerline{\bf Clifford V. Johnson\footnote{Also: Visiting Professor
    at the Centre for Particle Theory, Department of Mathematical
    Sciences, University of Durham, Durham DH1 3LE, England.}}

\bigskip
\bigskip

  \centerline{\it Department of Physics and Astronomy }
\centerline{\it University of
Southern California}
\centerline{\it Los Angeles, CA 90089-0484, U.S.A.}
\centerline{\small \tt johnson1@usc.edu}
\bigskip
\bigskip

\bigskip
\bigskip


\begin{abstract}

  Type 0A string theory in the $(2,4k)$ superconformal minimal model
  backgrounds and the bosonic string in the $(2,2k-1)$ conformal
  minimal models, while perturbatively identical in some regimes, may
  be distinguished non--perturbatively using double scaled matrix
  models. The resolvent of an associated Schr\"odinger operator plays
  three very important interconnected roles, which we explore
  perturbatively and non--perturbatively. On one hand, it acts as a
  source for placing D--branes and fluxes into the background, while
  on the other, it acts as a probe of the background, its first
  integral yielding the effective force on a scaled eigenvalue. We
  study this probe at disc, torus and annulus order in perturbation
  theory, in order to characterize the effects of D--branes and fluxes
  on the matrix eigenvalues.  On a third hand, the integrated
  resolvent forms a representation of a twisted boson in an associated
  conformal field theory. The entire content of the closed string
  theory can be expressed in terms of Virasoro constraints on the
  partition function, which is realized as wavefunction in a coherent
  state of the boson. Remarkably, the D--brane or flux background is
  simply prepared by acting with a vertex operator of the twisted
  boson. This generates a number of sharp examples of open--closed
  duality, both old and new.  We discuss whether the twisted boson
  conformal field theory can usefully be thought of as another
  holographic dual of the non--critical string theory.

\end{abstract}
\newpage \baselineskip=18pt \setcounter{footnote}{0}

\section{Introduction and Discussion}
\label{sec:introduction}

In recent years, it has become clear that the separation between open
and closed string descriptions of certain physical phenomena is very
much an artifact of perturbation theory. This means more than just
strong/weak coupling dualities linking different string theories, but
also includes transitions within a given theory from an open string
description ({\it e.g.}, involving D--branes) to a closed string one
({\it e.g.}, involving only fluxes), as the strength of a coupling
changes. Gauge/string dualities are the best known examples of this
type of phenomenon, and for that reason alone it is of interest to
learn more about how to describe them. For this, it is of course
necessary to go beyond perturbation theory, and furthermore, to
develop a far richer picture than that which is accessible using
only simple strong/weak coupling duality. We need to be able to define the
theories in question at arbitrary values of the couplings in order to
address the many subtle issues which can arise.

\begin{figure}[ht]
\begin{center}
  \includegraphics[scale=0.6]{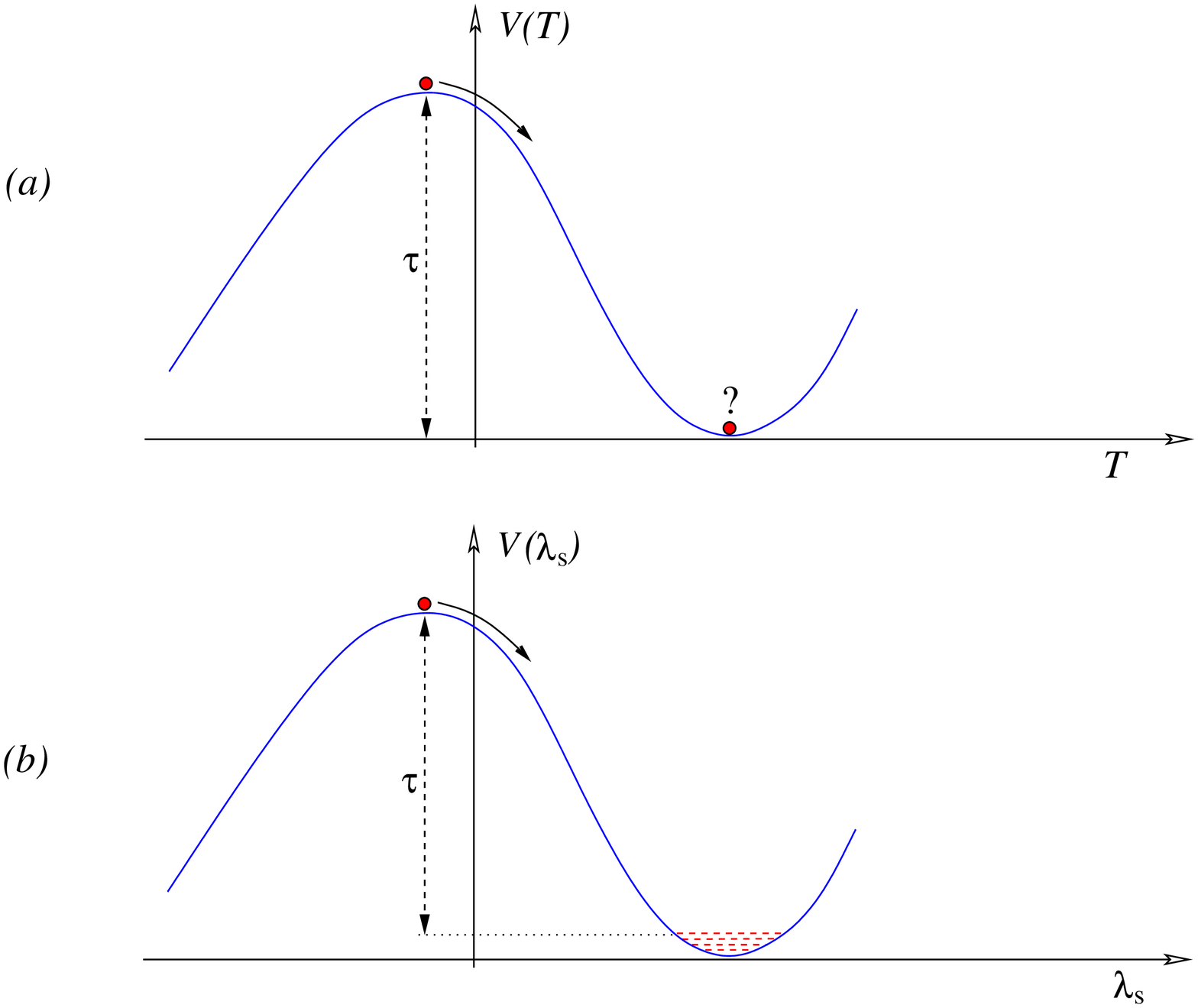}
\end{center}
\caption{\small The picture of tachyon condensation: {\it (a)} Incomplete in general, due to a lack of a description of the endstate as closed strings, and {\it (b)} Complete in the matrix model description, since the sea of eigenvalues at the minimum  is a holographic description of the closed string background.}
\label{fig:metaphor}
\end{figure}

Matrix models in the double scaling
limit\cite{Brezin:1990rb,Douglas:1990ve,Gross:1990vs} defined for us
the earliest examples of exactly solvable string theories with readily
accessible non--perturbative information. These types of models have
recently been interpreted as examples of open/closed dualities in a
very direct sense\cite{McGreevy:2003kb}, but also describe more subtle
of interesting open and closed string phenomena, as we shall recall.
In the direct sense, they represent certain decoupling limits of the
world--volume theories of $N$ D--branes in non--critical string
theories\cite{Fateev:2000ik,Teschner:2000md,Zamolodchikov:2001ah}, and
at large $N$ ---as in AdS/CFT--- they reconstruct their parent string
theories in specific spacetime backgrounds. These models are
particularly interesting in that they are not supersymmetric, the
D--branes are generically unstable, and as such their open string
modes are tachyonic. The matrix models in fact supply us with perfect
models of tachyon condensation\cite{Sen:1998tt,Sen:2002nu}, allowing
for the unified description of D--branes, closed strings, and the
transitions between
them\cite{Klebanov:2003km,McGreevy:2003ep,Alexandrov:2003nn,Gutperle:2003ij,Martinec:2003ka,Schomerus:2003vv}.
For example, the picture of D--brane decay as the process of rolling
down a tachyon potential to find a new minimum representing closed
strings always suffered somewhat from at least one conceptual
discomfort: Once one arrives at the bottom of the potential, one still
has the open string variables: Where are the closed strings? Without
the description of the closed string theory to which one is supposed to have
arrived, the picture drawn in figure~\ref{fig:metaphor}{\it (a)} is
really only a metaphor\footnote{See {\it e.g.}  ref.\cite{Sen:2003iv}
  for a discussion.}.
In the double scaled matrix model approach, such diagrams are refined
into more precise ones such as that depicted in
figure~\ref{fig:metaphor}{\it (b)}. The matrix model potential (after
double scaling) is the tachyon potential. In the eigenvalue
description, at large $N$ the eigenvalues form a droplet (or
collection of droplets) in one (or more) of the minima of this
potential. The collective excitations of the eigenvalue constituents
in these droplets are closed string physics, while the process of
moving an eigenvalue out of the collective and up to a extremum of the
potential represents the creation of a D--brane, whose tension is
given by the energy difference between the sea of collective
eigenvalues and the extremum.

Furthermore, it was shown some time ago how to describe, with matrix
models, string vacua containing a specific number (say, $\Gamma$) of
background D--branes, both for non--perturbatively unstable
models\cite{Kazakov:1990cq,Kostov:1990nf} and non--perturbatively
stable ones\cite{Dalley:1992br,Johnson:1994vk}. We will be examining
these definitions more in this paper. In each set of models, the
region of large positive cosmological constant, $\mu$, in each model
give identical perturbation theory, and can be identified as having
$\Gamma$ background D--branes.  The non--perturbatively stable models
allow one to continue to a second perturbative region, that of large
negative $\mu$, whose interpretation was unclear at the time the
models were first presented.  Recent work\cite{Klebanov:2003wg} has
shown that these non--perturbatively stable models are actually
descriptions of the type 0A string theories, and that the large
negative $\mu$ perturbative regime has an interpretation in terms of
there being $\Gamma$ units of R--R flux. This is an example of an
open/closed transition happening as a coupling ($\mu$) in the theory
changes. (See refs.\cite{Johnson:2003hy,Johnson:1992uy} for
generalizations to the rest of the minimal models for type~0A.)

In this paper we wish to examine in detail some of the key components
of these matrix model definitions, and sharpen the understanding of
many of the features of the models. A key role is played by the
diagonal of the resolvent ${\hat R}(z,\lambda_s)$ of the Schr\"odinger
operator ${\cal H}=-\partial^2_z+u(z)$, ($\lambda_s$ is a spectral
parameter) which enters in three logically distinct (but
interconnected) ways: {\it (a)} It acts as a source for placing
D--branes and fluxes into the background, as can be seen from the way
it enters the matrix model and the resulting string equations.  {\it
  (b)} It acts as a probe of the background, since its first
$z$--integral (after a Laplace transform) is the macroscopic loop
expectation\cite{Banks:1990df} operator. Its probe character also
follows from the fact that its first $z$--integral yields the
effective force on a scaled eigenvalue\cite{David:1990ge}. In order to
understand better the physics in the various regimes, we switch on the
background, corresponding to including the resolvent in the string
equations with strength $\Gamma$, and then we work in each
perturbative regime, using the integrated resolvent as a probe to see
the effects of the background. We study this probe at disc order in
large positive $\mu$ perturbation theory, and to annulus and torus
order in large negative $\mu$ perturbation theory, in order to
characterize the effects of D--branes and fluxes on the matrix
eigenvalues.

{\it (c)} The integrated resolvent forms a representation of a twisted
boson $\varphi(z)$ in an associated conformal field theory. The entire
content of the closed string theory can be expressed in terms of
Virasoro constraints\cite{Dijkgraaf:1991rs,Fukuma:1991jw} ---the
microscopic loop equations--- on the partition function which is
realized as wavefunction in a coherent state of the boson. The
creation modes of the boson are closed string operator insertions in
the string theory. The Virasoro constraints in the presence of
$\Gamma$ were worked out in ref.\cite{Dalley:1992br} and refined in
ref.\cite{Johnson:1994vk}.\footnote{See also ref.\cite{Itoh:1992vv},
  but note the comments on it in the discussion section of
  ref.\cite{Johnson:1994vk}.} It was observed in
ref.\cite{Johnson:1994vk} that the insertion of the D--brane (or flux)
background in this language corresponds to just acting with a vertex
operator\footnote{It appears that this fact was rediscovered many
  years later\cite{Fukuma:1999tj}.} of the twisted boson,
$V_\Gamma=:\exp(-\Gamma\varphi(\sigma)/2):$ The parameter $\sigma$
corresponds to the boundary cosmological constant.

This latter observation is particularly interesting since it makes
extremely explicit (if it were not manifestly so already) the fact
that in a fully non--perturbative framework, such as this one,
structures that we interpret as open strings in some perturbative
regime can be introduced in entirely closed string terms, built out of
a specific combination (derived in
refs.\cite{Itoh:1992vv,Johnson:1994vk}) of closed string
operators\footnote{This observation that one can move back and forth
  between closed and open string minimal model backgrounds by a
  redefinition of the closed string operators, explicitly at the level
  of the Virasoro constraints, was rediscovered
  recently\cite{Gaiotto:2003yb}.} . Here we see it succinctly in terms
of the vertex operator. This is of course one of the biggest lessons
to be taken away from all of this: Open/closed transitions and
strong/weak coupling dualities exist because the physics can be
specified independently of the perturbation theory (this is also the
central lesson of M--theory). When we can get a handle on the right
variables whose definition is not rooted in the world--sheet
expansions, we can describe the physics in a number of ways, involving
open or closed strings.

This also leads one to wonder whether the twisted boson framework
should be taken as an equivalent starting point for discussion of the
non--critical string theory in various backgrounds. Is there a sense
in which it is another (holographic) dual of the theory, or does it
rise above that, being the parent theory within which holography and
duality become manifest? This would go beyond the recent interesting
statements in the
literature\cite{Seiberg:2003nm,Kutasov:2004fg,Kazakov:2004du} that the
complex curve described by the analytically continued effective
potential is simply an alternative target space.  Instead, it would
suggest that the twisted boson should be given dynamical
meaning\footnote{We suspect that there is a relation with the work of
  ref.\cite{Aganagic:2003qj}, which recently discussed the
  organization of a number of topological string theory structures in
  terms of modes of a chiral boson $\phi$. Interestingly, in that
  work, of which we do not understand much (due to no fault of the
  authors) an exponentiation of the boson also plays a natural role in
  the insertion of D--branes in the background, which reminds us of
  the observation of ref.\cite{Johnson:1994vk}, discussed above.
  Clearly this connection deserves to be better understood.}.  Specific
solutions for the wavefunctions correspond to certain non--critical
string backgrounds we already know, and so we should expect that the
dynamics of $\varphi$ can give us more insight into non--perturbative
dynamics of string vacua, including perhaps time dependence, if we can
find analogous structures in higher dimensional string theories.

There is a fair amount of review material in this paper for which we
make no apology.  It sets up the language and conventions, and reminds
the reader of older (and often highly relevant) literature. The
following outline will help the well--informed reader avoid the parts
they wish to: In section~\ref{sec:stringequations}, we recall the
string equations for the bosonic and type~0A systems, noting that the
former are perturbatively equivalent to the latter in the large
positive cosmological constant regime. Section~\ref{sec:matrixbosonic}
is a detailed review of the double scaled matrix model origins of the
bosonic string equations, while section~\ref{sec:potentials} reviews
the details of how to extract the effective (tachyon) potential for
eigenvalues in the scaling limit. The non--perturbative effects of
D--branes (instantons) are recalled. Their tensions are computed from
the effective potentials for scaled eigenvalues, and are known from
explicit computations (at least for the unitary members of the
$(2,2k-1)$ series\cite{Alexandrov:2003nn}) to match the tensions of
the D--brane of the continuum theory.  The interpretation in terms of
D--brane world--volume theory and tachyon potentials is discussed in
detail there.  Sections~\ref{sec:matrixzeroay},~\ref{sec:matrixopen},
and~\ref{sec:potentialsagain} review and complete the derivation of
aspects of the doubled scaled matrix model origins of the type~0A
string equations, explains the connection to the bosonic models, and
discusses the differences in non--perturbative effects. This includes
a reminder of the role of the two seas of eigenvalues in stabilizing
the non--perturbative physics.  Section~\ref{sec:loopsbranes} turns to
backgrounds with non--trivial open string or R--R flux. The central
object in this section is the resolvent, which plays two connected
roles here.  The first is as a probe of the matrix model in the
language of macroscopic loops, or as the effective force on a scaled
eigenvalue. We re--derive the effective (tachyon) potential of
section~\ref{sec:potentials} from this point of view, now including
terms higher order in perturbation theory. The second role is as a
source in the string equations which inserts a number of D--branes, or
units of flux, into the background.  This section has both of these
roles working in tandem, since we study the loop operator (eigenvalue
dynamics) in the presence of the non--trivial backgrounds.  So after a
brief review of the properties of the string equations for these
situations, we remind the reader of the properties of the scaled loop
operator in these theories, and explain the systematics of developing
the results for the operator perturbatively, using the string
equations together with the Gel'fand--Dikii equation. We derive
compact expressions for the loop operator and hence the eigenvalue
distributions and effective potentials to disc order (and to annulus
and torus order), and analyze the results. In
section~\ref{sec:loopsbranesagain} we remark upon the fluidity of the
formalism in its treatment of the closed string and open string
sectors. We see that the open string (or flux) backgrounds can be
implemented in terms of a specific preparation of closed string
operators. This is made much more transparent ---and suggestive--- in
an associated conformal field theory language, which organizes
everything. We remind the reader of the
formalism\cite{Dijkgraaf:1991rs,Fukuma:1991jw} of the $\IZ_2$ twisted
boson, and of how the entire content of the string theory is expressed
as Virasoro constraints on the partition function. This partition
function is realized in the twisted boson framework as a coherent
state. We then discuss how\cite{Johnson:1994vk} the open string or
flux background is simply realized by placing a vertex operator into
the coherent state background. This shows the remarkable transparency
of the formalism with regards working with open or closed string
backgrounds.

\newpage

\section{The String Equations}
\label{sec:stringequations}
Consider the following equations:
\begin{equation}
{\cal R}=2\nu\Gamma {\hat R}(z,\sigma)\ ,
  \labell{eq:nonpertbos}
\end{equation}
and
\begin{equation}
(u-\sigma){\cal R}^2-\frac{1}{2}{\cal R}{\cal R}^{''}+\frac{1}{4}({\cal R}^{'})^2
  =\nu^2\Gamma^2\ .\labell{eq:nonpert}
\end{equation}
Here $u(z)$ is a real function of the real variable $z$, a prime
denotes $\nu \partial/\partial z$, where $\nu$ is a constant, and
$\sigma$ and $\Gamma$ are parameters we shall discuss further shortly.
We define the quantity:
\begin{equation}
  \label{eq:R}
{\cal
  R}=\sum_{k=0}^\infty\left(k+\frac{1}{2}\right)R_k\ ,   
\end{equation}
where the $R_k$ ($k=0,\ldots$)
 are polynomials in $u(z)$ and its $z$--derivatives. They
are related by a recursion relation:
\begin{equation}
  \labell{eq:recursion}
  R^{'}_{k+1}=\frac{1}{4}R^{'''}_k-uR^{'}_k-\frac{1}{2}u^{\prime}R_k\ ,
\end{equation}
and are fixed by the constant $R_0$, and the requirement that the rest
vanish for vanishing $u$. The first few are:
\begin{equation}
  \labell{eq:firstfew}
  R_0=\frac{1}{2}\ ;\quad R_1=-\frac{1}{4}u\ ;\quad R_2=\frac{1}{16}(3u^2-u^{''})\ .
\end{equation}
The $k$th model is chosen by setting all the other $t$s to zero except
$t_0\equiv z$, and $t_k$, the latter being fixed to a numerical value
such that ${\cal R}={\cal D}_k-z$. The ${\cal D}_k$ are normalized
such that the coefficient of $u^k$ is unity, {\it e.g.}:
\begin{eqnarray}
{\cal D}_1=u\ ,\quad {\cal D}_2=u^2-\frac{1}{3}u^{''}\ ,\quad {\cal D}_3=u^3- u u^{''}-\frac{1}{2}(u^{'})^2+\frac{1}{10}u^{''''}\ .
  \labell{eq:diffpolys}
\end{eqnarray}
The quantity 
\begin{equation}
  \labell{eq:resolvent}
{\hat R}(z,\lambda)=<z|\frac{1}{({\cal H}-\lambda)}|z>\ ,  \quad {\rm where}\quad {\cal H}=-\nu^2\frac{\partial^2}{\partial z^2}+u(z)\ ,
\end{equation}
is the diagonal of the resolvent of the Hamiltonian ${\cal H}$ and it
satisfies the Gel'fand--Dikii
equation\cite{Gelfand:1976A,Gelfand:1976B,Gelfand:1975rn}:
\begin{equation}
  \labell{eq:resolventequation}
  4(u-\lambda){\hat R}^2-2{\hat R}{\hat R}^{''}+({\hat R}^{'})^2
  =1\ .
\end{equation}
The function $u$ defines the partition function $Z=\exp(-F)$ of a
string theory {\it via}:
\begin{equation}
u(z)=\nu^2\frac{\partial^2 F}{\partial \mu^2}\Biggl|_{\mu=z}\ ,
  \labell{eq:partfun}
\end{equation}
where $\mu$ is the coefficient of the lowest dimension operator in
the world--sheet theory. It is the world--sheet cosmological constant
in the case $k=2$ for the bosonic string (in which case we take
equation~\reef{eq:nonpertbos}) and $k=1$ for the type 0A string, in
which case we take equation~\reef{eq:nonpert}). For the case $k=1$ for
the bosonic string, $\mu$ couples to a boundary operator measuring
lengths.

The equations~\reef{eq:nonpertbos} and~\reef{eq:nonpert} have exactly
the same physics for large $z$ (equivalent to $\mu$; we will sometimes
use the two interchangeably), which corresponds to a perturbative
regime. Let us set $\Gamma=0$ and $\sigma=0$ for now. For example, for
$k=2$, we have for the free energy (discarding non--universal terms):
\begin{equation}
F=\frac{4}{15}g_s^{-2}+\frac{1}{24}\log\mu-\frac{7}{1440}g_s^2+\cdots\ ,
  \labell{eq:partfunexpand}
\end{equation}
where the dimensionless parameter $g_s={\nu}{\mu^{-5/4}}$ is the
closed string coupling. We see the expansion in terms of closed string
worldsheets of topology $\chi=2-2h$ ($h$ is the number of handles),
each coming with factor $g_s^{-\chi}$. The dependence on the sphere
displays bosonic KPZ scaling $F(\mu)\sim\mu^{2-\gamma_{str}}$ for the
critical exponent (string susceptibility) $\gamma_{\rm str}=-1/k$,
and $k=2$ here. For $k=1$, $Z\sim \mu^3$, which should be discarded
with other non--universal physics, and so this theory has no
contribution from any closed string world--sheets. It is a topological
theory, as we will recall below. The closed string coupling for this $k=2$
case is $g_s= \nu \mu^{-3/2}$. For general $k$, it is:
\begin{equation}
  \labell{eq:closedstringcoupling}
  g_s=\frac{\nu}{\mu^{1+\frac{1}{2k}}}\ ,
\end{equation}
which fits with the fact that 
\begin{equation}
  \labell{eq:udepends}
  u(z)=z^{\frac{1}{k}}+\cdots
\end{equation}
and hence:
\begin{equation}
  \labell{eq:free}
  F\simeq\frac{\mu^{\frac{1}{k}+2}}{\nu^2}\ .
\end{equation}
Non--perturbatively, the situations for each equation, and hence the
models they describe, are different. For the case of
equation~\reef{eq:nonpertbos}, there is a well--known
instability\cite{Brezin:1990rb,Douglas:1990ve,Gross:1990vs}. The
physics is non--perturbatively problematic for even $k$, there being
no sensible real solutions for $u(z)$ which match onto the large $z$
perturbation theory just discussed. This fits with the fact that the
model is rendered unstable through a decay channel mediated by
instanton effects\cite{David:1990ge,Dalley:1991zr} which we now think
of as the nucleation of D--branes.

Let us turn to equation~\reef{eq:nonpert}, starting again with
$\sigma$ and $\Gamma$ set to zero. This equation enjoys better
non--perturbative properties than the original string
equation~\reef{eq:nonpertbos}, (see
refs.\cite{Dalley:1992qg,Dalley:1992vr,Dalley:1992yi,Dalley:1992br,Johnson:1992pu,Johnson:thesis})
while sharing the same perturbative physics for large positive $z$, as
follows from the fact that ${\cal R}=0$ is also a solution of this
equation, and that for large $|z|$, the leading order equation is
$u(u^k-z)=0$. So the behaviour $u=z^{\frac{1}{k}}+\hdots$ can be
chosen for large positive $z$, and the perturbative expansion is
exactly the same as that given in equation~\reef{eq:partfunexpand}.

\begin{figure}[ht]
\begin{center}
  \includegraphics[scale=0.55]{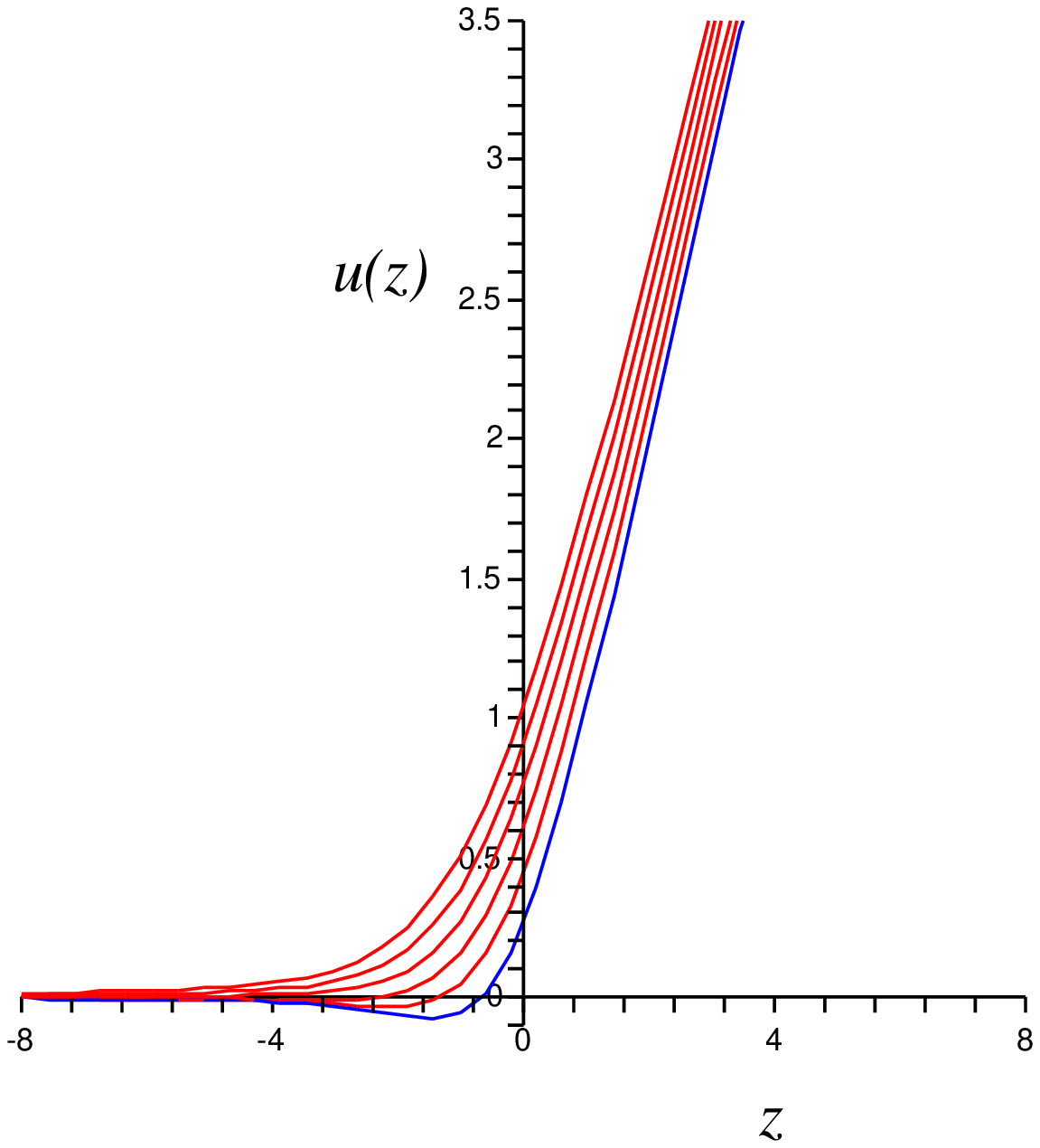} \includegraphics[scale=0.55]{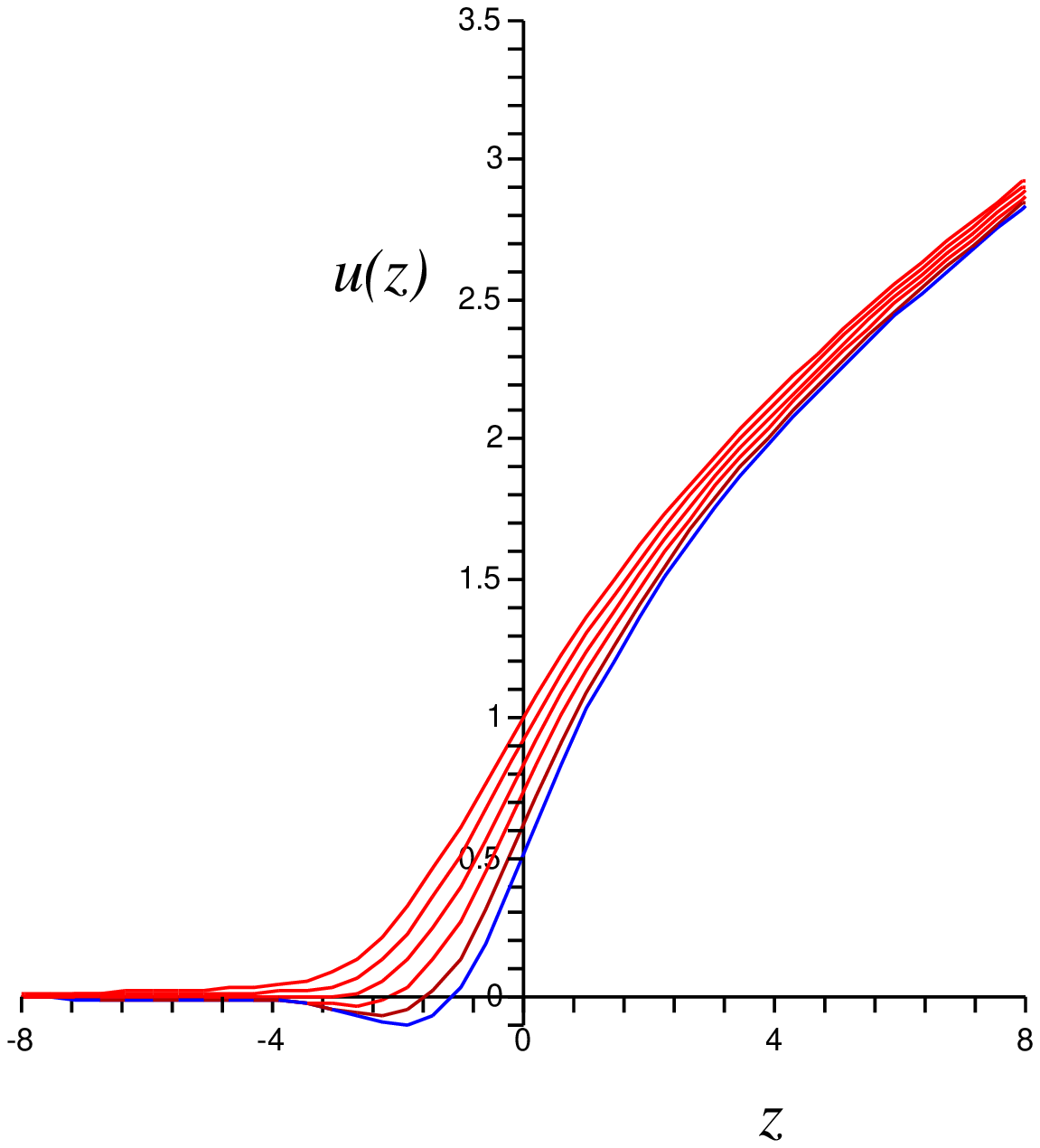}
\end{center}
\caption{\small Numerical solutions to equation~\reef{eq:nonpert} for $u(z)$: {\it (a)} The case $k=1$, and {\it (b)} the case $k=2$, for a range of values of  the parameter $\Gamma$, with $\sigma=0$. The very bottom curve in each case is the case of $\Gamma=0$. The solutions all asymptote to $u=z^{1/k}$ for large positive $z$ and $u=0$ for large negative $z$.} 
\label{fig:gammaplots}
\end{figure}

For large negative $z$, however, we can choose a completely different
behaviour for solutions of equation~\reef{eq:nonpert} from that
available for equation~\reef{eq:nonpertbos}. We can choose the
asymptotic $u=0$ in that regime, having the physics on the sphere
vanish for all $k$. The next non--vanishing order is of a universal
form for all $k$, and is:
\begin{equation}
  \labell{eq:torus}
  u(z)=-\frac{1}{4}\frac{\nu^2}{z^2}+\cdots\ ,
\end{equation}
yielding a universal torus contribution for all models.  The full
solutions of the equation with these asymptotics for large $\pm z$,
for all $k$, are believed to be unique and pole--free supplying a
fully consistent non--perturbative physical string theory. See for
example the bottom curves of the families depicted in
figure~\ref{fig:gammaplots}{\it (a)} and {\it (b)}.

The cases of $k=1,2$ and 3 have been studied extensively, and (at
least) $k=1$ is known to exist uniquely (since it
maps\cite{Morris:1990bw,Morris:1992zr} to the case of the simplest
Unitary matrix model critical
point\cite{Periwal:1990gf,Periwal:1990qb} which defines unique
physics\cite{Crnkovic:1990ms,Watterstam:1990qs}) and this at least
implies a unique family for all $k$ connected by KdV
flows\cite{Johnson:1992pu}.

These solutions were proposed as non--perturbative completions of the
sick bosonic string models a while
ago\cite{Dalley:1992qg,Dalley:1992vr,Dalley:1992yi,Dalley:1992br,Johnson:1992pu,Johnson:thesis}.
That statement has now been refined by recent
work\cite{Klebanov:2003wg}: The models have an interpretation as type
0A models.  This is consistent with a number of facts, among which is
the fact\cite{Klebanov:2003wg} that the spectrum of operators of the
Liouville dressed $(2,2k-1)$ conformal minimal models match exactly
those of the super--Liouville dressed $(2,4k)$ superconformal minimal
models.

\section{Matrix Models and Double Scaling 1: Bosonic Strings}
\label{sec:matrixbosonic}

Both sets of models can be derived from one--matrix models in the
double scaling limit and we can choose Hermitian matrices if we so
wish.  Equation~\reef{eq:nonpertbos} for $\Gamma=0$ arose for
Hermitian matrix models, which can be written in the form:
\begin{equation}
{\cal Z}=\int dM \exp\left\{-\frac{N}{\gamma}{\rm \Tr}\left(V(M)\right)\right\}\ ,
  \labell{eq:hermitian}
\end{equation}
where $M$ is an $N{\times}N$ Hermitian matrix, and the coefficients of
the polynomial potential $V(M)$ and the parameter $\gamma$ are tuned
to critical values as $N\to\infty$.  Recall that the Feynmann diagrams
of the matrix model are interpreted as dual to discretisations of the
string worldsheets, organized in a genus expansion for large $N$. The
path integral is then to be thought of as a particular regularization
which enumerates the sum over surfaces. The typical size of a cell in
the discretization is set by a parameter we shall call $\delta$. The
critical potentials are those which pick out surfaces with diverging
area, so that in the continuum limit $\delta=0$ these give finite area
contributions to the path integral.

The typical method of study\cite{Brezin:1978sv} is to diagonalize $M$
using a unitary matrix $U$ into the form $M=U \Lambda U^\dagger$,
where $\Lambda={\rm
  diag}\left\{\lambda_1,\lambda_2,\ldots,\lambda_N\right\}$. There is
a Jacobian for this change of variables, which is the square of the
Van der Monde determinant $\prod_{i<j}(\lambda_i-\lambda_j)$, and the
resulting model is (after dividing by a normalization factor ${\rm
  Vol}(U(N))$):
\begin{equation}
  \labell{eq:party}
{\cal Z}=\int\prod_i d\lambda_i \prod_{i\neq j}(\lambda_i-\lambda_j)^2\exp\left\{-\frac{N}{\gamma}\sum_{i=1}^NV(\lambda_i)\right\}\ ,  
\end{equation}

An efficient way of working\cite{Bessis:1980ss} is to use polynomials
$P_n(\lambda)\sim\lambda^n+\cdots$ which are orthogonal with respect
to the measure $d\mu(\lambda)=d\lambda e^{-N/\gamma V(\lambda)}$,
\begin{equation}
\int P_n(\lambda)P_m(\lambda) d\mu(\lambda)=h_n\delta_{nm}\ .
  \labell{eq:polys}
\end{equation}
The polynomial $\lambda P_n(\lambda)$ must be equal to $P_{n+1}$ plus
a correction, which for even potentials (on which we will focus here)
is simply:
\begin{equation}
\lambda P_n(\lambda)=P_{n+1}(\lambda)+R_n P_{n-1}(\lambda)\ ,
  \labell{eq:relate}
\end{equation}
where we have defined the coefficient $R_n$. Using the orthogonality condition it is given in terms of the $h_n$s as $R_{n+1}=h_{n+1}/h_n$.

After a little more work, one can write the vacuum energy of the model
as\cite{Bessis:1980ss}:
\begin{equation}
  \labell{eq:freetoo}
F=-\sum_{n=1}^N (N-n)\log R_n+\cdots\ ,
\end{equation}
where the ellipsis denote parts which are non--universal, and the
entire problem is one of determining the $R_n$s.

The content of the matrix model can be extracted from the relation:
\begin{equation}
  \label{eq:identity}
  \int d\mu(\lambda) P_{n-1}(\lambda)\frac{d}{d\lambda}P_n(\lambda)=nh_n\ ,
\end{equation}
which amounts, after integrating by parts, to a compact master
equation:
\begin{equation}
  \labell{eq:master}
 \sqrt{R_n} <n-1|V^{'}({\hat \lambda})|n>=\frac{\gamma n}{N}\ ,
\end{equation}
where we have used a bra--ket representation of the orthogonal
polynomials and the integration over product pairs of them with respect to the
measure. Functions integrated with respect to the measure become
operator valued in this representation:
\begin{eqnarray}
|n>=\frac{P_n}{\sqrt{h_n}}\ ,\quad {\hat n}|n>=n\  ,\quad {\cal A}|n>=|n-1>\ ,\quad {\cal A}^\dagger|n>=|n+1>\nonumber\\
<m|n>=\delta_{mn}\ ,\quad {\cal A}R_{\hat n}=R_{{\hat n}+1}{\cal A}\ , \quad {\cal A}^\dagger R_{\hat n}=R_{{\hat n}-1}{\cal A}^\dagger\ .
  \labell{eq:operators}
\end{eqnarray}
The recursion relation~\reef{eq:relate} define 
\begin{equation}
  \labell{eq:operate}
  {\hat \lambda}=\sqrt{R_{\hat n}}{\cal A}^\dagger+{\cal A}\sqrt{R_{\hat n}}\ ,
\end{equation}
which allows a rewriting of the master equation in the equivalent form:
\begin{equation}
  \label{eq:masteralt}
  <n|{\hat \lambda}V^{\prime}({\hat\lambda})|n>=\frac{\gamma}{N}(2n+1)\ ,
\end{equation}
which will be useful later on.  It is convenient to write our master
equation~\reef{eq:master} a touch more compactly as:
\begin{equation}
  \labell{eq:mastermore}
  <n|{\cal A}^\dagger V^{'}({\hat \phi})|n>=\frac{\gamma n}{N}\ ,
\end{equation}
where
\begin{equation}
  \labell{eq:operatorchange}
  {\hat \phi}={\cal A}^\dagger+{\cal A}R_{\hat n}\ ,\quad \sqrt{h_{\hat n}}{\hat \phi}={\hat \lambda}\sqrt{h_{\hat n}}\ .
\end{equation}
The master equation is now actually a recursion relation for the
$R_n$, as can be seen by expanding for a given potential $V(\lambda)$.
The higher the order of $V(\lambda)$, the more terms there are in the recursion
relation.

The continuum limit is then easily approached in these variables. At
large $N$ we pass to a continuum limit in eigenvalues space as well,
and write $x=n/N$, $R(x)=R(n/N)$, $\epsilon=1/N$. The master equation
is:
\begin{equation}
  \labell{eq:mastercontinuum}
  <x|{\cal A}^\dagger V^\prime({\hat \phi})|x>=\gamma x\ ,
\end{equation}
and the free energy is:
\begin{equation}
  \labell{eq:freecontinuum}
  F=-N^2\int_0^1 dx(1-x)\log R(x)\ .
\end{equation}
Focusing on the sphere contribution means dropping terms subleading
in $N$ and so the difference equations become polynomial relations in
$R(x)$:
\begin{equation}
  \labell{eq:polypol}
1-  \gamma x = P(R(x))\ .
\end{equation}
The required divergent behaviour of the free energy $F\sim
(\gamma_c-\gamma)^{2-1/k}$ occurs when the polynomial in $R(x)$
acquires $k$ multiple roots: $P(R)=(R-R_c)^k$. This amounts to
choosing particular values of the couplings in the
potential\cite{Kazakov:1989bc} and also tuning $\gamma$ to a critical
value $\gamma_c$. The critical value of $R(x)$ at this point can be
chosen to be equal to 2, and in the neighbourhood $\gamma x=1$,
without loss of generality. We can choose a potential of degree $2k$
to get the above behaviour, and a convenient basis
is\cite{Gross:1990aw}:
\begin{eqnarray}
V_1(\phi)&=&\frac{1}{4}\phi^2\ ;\nonumber\\
V_2(\phi)&=&\frac{1}{2}\phi^2-\frac{1}{48}\phi^4\ ;\nonumber\\
\vdots &=&\vdots\nonumber\\
V_k(\phi)&=&\sum_{i=1}^k (-1)^{i-1}\frac{k!(i-1)!}{2^i(k-i)!(2i)!} \phi^{2i}\ .
  \labell{eq:criticalpots}
\end{eqnarray}
 
Finite physics is extracted by allowing $R(x)$ and $\gamma x$ to
approach their critical values at the right rate as $\delta\to 0$. The
quantity $\epsilon =1/N$ must also be scaled with $\delta$ so as to
approach zero at the correct rate:
\begin{eqnarray}
\gamma=1-\frac{\mu}{4}\delta^{2k}\ ;\nonumber\\
\gamma x=1-\frac{z}{4}\delta^{2k}\ ; \nonumber\\
R(x)=2-u(z) \delta^2\ ; \nonumber\\
\epsilon=\frac{\nu}{4\sqrt{2}}\delta^{2k+1}\ .
  \labell{eq:critical}
\end{eqnarray}
(Numerical prefactors of the scaled quantities $\mu,z,u(z)$ and $\nu$
are chosen for convenience.)  To go beyond the sphere one simply keeps
the $1/N$ contributions in the master difference equation, and in the
large $N$ limit it becomes a differential equation.  Inserting the
scalings into the master equation and expanding about $\delta=0$ gives
for the first non--vanishing terms, a differential equation of order
$2k-2$ at order $\delta^{2k}$. So in taking the limit $\delta\to0$,
this is the surviving physics and it is in fact
equation~\reef{eq:nonpertbos}, with $\Gamma=0$, $\sigma=0$ and all
$t$s set to zero except $t_0\sim z$ and $t_k$.

By introducing dimensionful coefficients $t_k$ for each potential, we
can consider the general model. From the point of view of the $k$th
theory, the other $t_k$s represent coupling to closed string operators
${\cal O}_k$. It is well known that the insertion of each operator can
be expressed in terms of the KdV
flows\cite{Douglas:1990dd,Banks:1990df}:
\begin{equation}
  \labell{eq:kdvflows}
  \frac{\partial u}{\partial t_k}= R^{'}_{k+1}\ .
\end{equation}
The operator ${\cal O}_0$ couples to $t_0$, which is in fact $-4z$, the
cosmological constant (in the unitary model). So ${\cal O}_0$ is often
referred to as the puncture operator, which yields the area of a
surface by fixing a point which is then integrated over in the path
integral.  The function $u$ itself can be thought of as the two point
function of the puncture operator. 

\section{Effective Tachyon Potentials and D--Branes: Bosonic Strings}
\label{sec:potentials}
It is also important to keep in mind the eigenvalue picture. The
eigenvalues take values on the real line. The Van der Monde
determinant acts as a repulsive potential, driving them apart, but
they are confined by the potential $V(\lambda)$ to form a droplet of
finite size at large $N$.  At large $N$, at the spherical level, one
can solve for a density $\rho(\lambda)$ of eigenvalues on the line. A
relation between the orthogonal polynomial quantity $R(x)$ and the
eigenvalue density $\rho(\lambda)$ can be derived at sphere
level\cite{Bessis:1980ss}:
\begin{equation}
  \labell{eq:integralrep}
\rho(\lambda)=\int_0^1\frac{dx}{\pi}\frac{\theta(4R(x)-\lambda^2)}{\sqrt{4R(x)-\lambda^2}}\ .
\end{equation}
From this expression, it is clear that the ends of the eigenvalue
density are located at $\lambda_c=\pm2\sqrt{R_c}$. Recalling that the
neighbourhood of $x=1$ is where the interesting physics comes from,
we must pick one of them to which to scale our physics:
\begin{equation}
  \labell{eq:scaletheend}
  \lambda=2\sqrt{2}-\frac{\lambda_s}{\sqrt{2}}\delta^2\ .
\end{equation}
This, together with the scalings~\reef{eq:critical} gives an
expression for the scaled density:
\begin{equation}
  \labell{eq:scaleddensity}
  \rho(\lambda_s)=-\frac{\delta^{2k-1}}{16\pi}\int^\mu\frac{dz}{\sqrt{\lambda_s-u(z)}}=\frac{\delta^{2k-1}}{16\pi}\int_{\mu^{1/k}}^{\lambda_s}\frac{k u^{k-1}du}{\sqrt{\lambda_s-u(z)}}\ ,
\end{equation}
where we have used that $u(z)=z^{1/k}$ on the sphere.  Explicit
computation of the integral yields\cite{Dalley:1992vr}:
\begin{equation}
  \labell{eq:densityk}
  \rho_{(k)}(\lambda_s)=\frac{k!\delta^{2k-1}}{16\pi}\sum_{n=1}^k\frac{2^n}{(k-n)!(2n-1)!!}(\lambda_s-\mu^\frac{1}{k})^{n-\frac{1}{2}}\mu^{1-\frac{n}{k}}+\cdots\ ,
\end{equation}
where we remind the reader again that this is valid only at the sphere
level.

Crucially, note that since there is a common factor
$(\lambda_s-\mu^{1/k})^{\frac{1}{2}}$, the eigenvalue density vanishes
at $\lambda_s=\mu^{1/k}$ as $\lambda^{\frac{1}{2}}$. There are $k-1$
more zeros which are located away from this endpoint. For $k$ even
there is another zero at $\lambda<0$ and $k-2$ zeros off the real line
in pairs. For $k$ odd, there are no other zeros on the real line, and
only pairs of zeros off the real line. In general, $\rho(\lambda)$ is real on
the part of the real line where the eigenvalue are located but its
complex parts are meaningful also, as we shall see.

\begin{figure}[ht]
\begin{center}
  \includegraphics[scale=0.55]{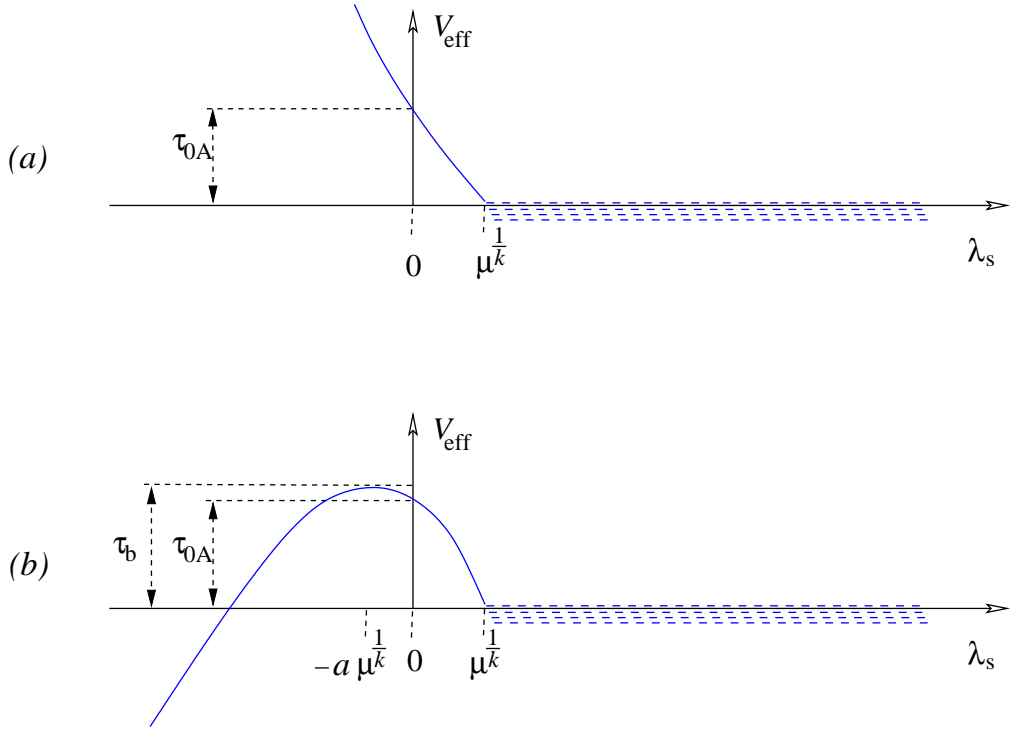}
\end{center}
\caption{\small The effective potential for one eigenvalue in the scaling limit, computed at string theory tree level. This is equivalent to the tachyon potential on the world--volume theory of the D--branes. The sea of eigenvalues is to the right in each sketch,  on which $V_{\rm eff}=0$. The collective physics of the sea describes closed string physics.  A D--brane is described by a single eigenvalue separated from the sea, and its decay by rolling back of the eigenvalue into the sea.  The energy released in decay sets the tension of the D--brane. Shown is the difference between the D--brane tension in the bosonic ($\tau_{\rm b}$) and  that of the type 0A theory ($\tau_{\rm 0A}$) for the cases of {\it (a)} $k$ odd   and  {\it (b)} $k$ even. In the 0A case, the potential stops at $\lambda_{\rm s}=0$, setting a lower value for the D--brane tension. }
\label{fig:effpots}
\end{figure}

A most important quantity is the effective potential that an
individual eigenvalue sees\cite{David:1990ge}. It is minus the change
in energy as the eigenvalue moves from $\lambda_i$ to $\lambda_f$, and
is given by the expression:
\begin{equation}
  \labell{eq:effpot}
  V_{\rm eff}(\lambda)= N\int_{\lambda_f}^{\lambda_i}d\lambda (V^{'}(\lambda)-2F(\lambda))=2\pi N {\rm Im}\int_{\lambda_f}^{\lambda_i}d\lambda
  \rho(\lambda)\ ,
\end{equation}
where 
\begin{equation}
F(\lambda)=\int \frac{\rho(\xi)d\xi}{\lambda-\xi}
  \labell{eq:resolventagain}
\end{equation}
is the matrix model resolvent.  So starting with the eigenvalue on the
cut and moving it to anywhere else on the cut costs no energy, since
$\rho(\lambda)$ is real there.  In scaled variables the effective
potential becomes\cite{Dalley:1992vr}:
\begin{equation}
  \labell{eq:veffscaled}
  V_{{\rm
  eff}}^{(k)}(\lambda_s)=\frac{2\pi}{\nu\delta^{2k+1}}\int_{\lambda_s}^{\mu^{1/k}}\delta^2\rho(\lambda^{'}_s)d\lambda^{'}_s\
  ,\qquad\lambda_s<\mu^{\frac{1}{k}}\ .
\end{equation}
Inserting our previous expression for the density
gives\cite{Dalley:1992vr}:
\begin{equation}
  \labell{eq:effectivepotential}
  V_{{\rm
  eff}}^{(k)}(\lambda_s)=\frac{1}{\nu}\sum_{n=1}^k\frac{k!(-2)^{n+1}}{(k-n)!(2n+1)!!}(\mu^{\frac{1}{k}}-\lambda_s)^{n+\frac{1}{2}}\mu^{1-\frac{n}{k}}+\cdots\
  ,\qquad\lambda_s<\mu^{\frac{1}{k}}\ .
\end{equation}
Looking at the first two cases explicitly is illustrative:
\begin{eqnarray}
V_{\rm eff}^{(1)}(\lambda_s)&=&\frac{4}{3\nu}(\mu-\lambda_s)^\frac{3}{2}+\cdots\ ,\quad \lambda_s<\mu\ ,
\nonumber\\
V_{\rm
  eff}^{(2)}(\lambda_s)&=&\frac{8}{3\nu}(\mu^{\frac{1}{2}}-\lambda_s)^\frac{3}{2}\mu^{\frac{1}{2}}-\frac{16}{15\nu}(\mu^{\frac{1}{2}}-\lambda_s)^\frac{5}{2}+\cdots\ ,\quad \lambda_s<\mu\ . \labell{eq:effpotonetwo}
\end{eqnarray}
Their behaviours are sketched in figure~\ref{fig:effpots}. 

The key point is that there is a turning point of the effective
potential at every zero of the eigenvalue density. In particular we
see that the even $k$ cases have a maximum of the potential at one
point on the real axis at $\lambda_s<0$. The odd $k$ cases have an
effective potential which forever rises to the left. To the right in
figure~\ref{fig:effpots} is the sea of the other eigenvalues, starting
at $\lambda_s=\mu^{\frac{1}{k}}$ and going off to infinity.

In the spirit of ref.\cite{McGreevy:2003kb}, the interpretation of
this physics in the continuum string theory is as follows. The matrix
model, once double--scaled, is the theory on the world--volume of $N$
unstable D--branes in the continuum string theory. This model is
constituted by a matrix--valued tachyonic field made by open strings
interconnecting the $N$ D--branes. The computation above is a precise
determination of the potential of the tachyon field at sphere level in
string perturbation theory. We have a precise model in which we see
holography at work, together with a model of tachyon condensation. To
the right is the sea of $N$ eigenvalues. The collective dynamics of
these represents closed string physics: Bosonic non--critical string
theory given by Liouville theory coupled to the $(2k-1,2)$ conformal
minimal models. For this model, the full family of critical potentials
given in equation~\reef{eq:criticalpots} forms a basis for the closed
string deformations. They can be added to the model with coefficients
$t_k$ and are dual to inserting closed string operators ${\cal O}_k$
to the model. These are the Liouville dressed conformal field theory
operators. As is well known\cite{Douglas:1990dd,Banks:1990df}, this
family of operators is organized by the KdV flows of
equation~\reef{eq:kdvflows}. 

There is more, however\cite{McGreevy:2003kb}. A single eigenvalue can
be removed from the sea and moved up the potential. This describes the
creation of a D--brane of the continuum string theory, and it can
decay back to closed strings by rolling back down the potential to
rejoin the sea.  Consistent with this identification is the
observation of the behaviour of the tension of the D--brane. The
tension of the D--brane should be the energy needed to move up the
potential to a definite position which goes like: $\lambda_s\sim
c\mu^{\frac{1}{k}}$, where $c$ is a constant. Upon examination of the
potential $V_{\rm eff}^{(k)}$ it can be seen to behave as:
\begin{equation}
  \labell{eq:tensiongeneral}
  \tau\sim V_{\rm eff}^{(k)}(c\mu^{\frac{1}{k}})\sim C\frac{\mu^{1+\frac{1}{2k}}}{\nu}=\frac{C}{g_s}\ ,
\end{equation} where $C$ is a constant.
This is the correct behaviour for a D--brane's tension.

For $k$ even, there is a special class of D--branes formed by placing
an eigenvalue at the maximum of the potential, located at
$\lambda_s=\lambda_c$. This is a D--brane which dominates the
non--perturbative contributions to the closed string physics, an
observation going back to ref.\cite{Shenker:1990uf}. Its tension is:
\begin{equation}
  \labell{eq:tensionspecial}
  \tau^{(k)}_{\rm b}\sim V_{\rm eff}^{(k)}(\lambda_c)\ ,\qquad {\rm for }\,\,\,k \quad {\rm even}\ .
\end{equation}
For example, the case $k=2$, since the potential has a maximum at
$\lambda_c=-\frac{1}{2}\mu^{\frac{1}{2}}$, we
get\cite{Shenker:1990uf,David:1991sk}:
\begin{equation}
  \labell{eq:tensionktwo}
  \tau^{(2)}_{\rm b}=V_{\rm eff}^{(2)}\left(-\frac{1}{2}\mu^{\frac{1}{2}}\right)=\frac{4\sqrt{6}}{5}\frac{1}{g_s}\ ,\quad g_s=\frac{\nu}{\mu^{\frac{5}{4}}}\ .
\end{equation}
This actually shows up as the action of the D--instanton effects
arising by studying the leading contribution to effects invisible in
perturbation theory. Taking a solution $u_0(z)$ of the equation:
\begin{equation}
  \labell{eq:painleveI}
  -\frac{1}{3}u_0^{''}+u_0^2=z\ ,
\end{equation}
one can consider a perturbation $\epsilon(z)$, and study another
solution $u(z)=u_0(z)+\epsilon(z)$. Assuming that $\epsilon$ is small,
and of the form $\epsilon=\exp(-f)$, where $f>>f'>>f''$, {\it etc.},
the result\cite{Shenker:1990uf} is for $u_0=z^{1/2}$:
\begin{equation}
  \labell{eq:instanton}
  \epsilon=e^{-\frac{4\sqrt{6}}{5}\frac{1}{g_s}}\ .
\end{equation}
This D--brane, (and its analogue for all $k$ even) as can be seen from
the effective potential, destabilizes the theory for $k$ even, since
the eigenvalue can roll off to infinity to the left of the
figure~\ref{fig:effpots}{\it (b)}. This process dooms the $k$ even
models\cite{David:1990ge,Dalley:1991zr}. Note that we should be
careful not to confuse the instanton effects derived from analyzing
perturbation theory and those of the real destabilising instanton see
in the effective potential. They match in this case, but existence of
one does not imply the other, as we shall recall for the type~0A
models\cite{Dalley:1992vr}.

\section{Matrix Models and Double Scaling II: Type 0A Strings}
\label{sec:matrixzeroay}
\begin{figure}[ht]
\begin{center}
  \includegraphics[scale=0.5]{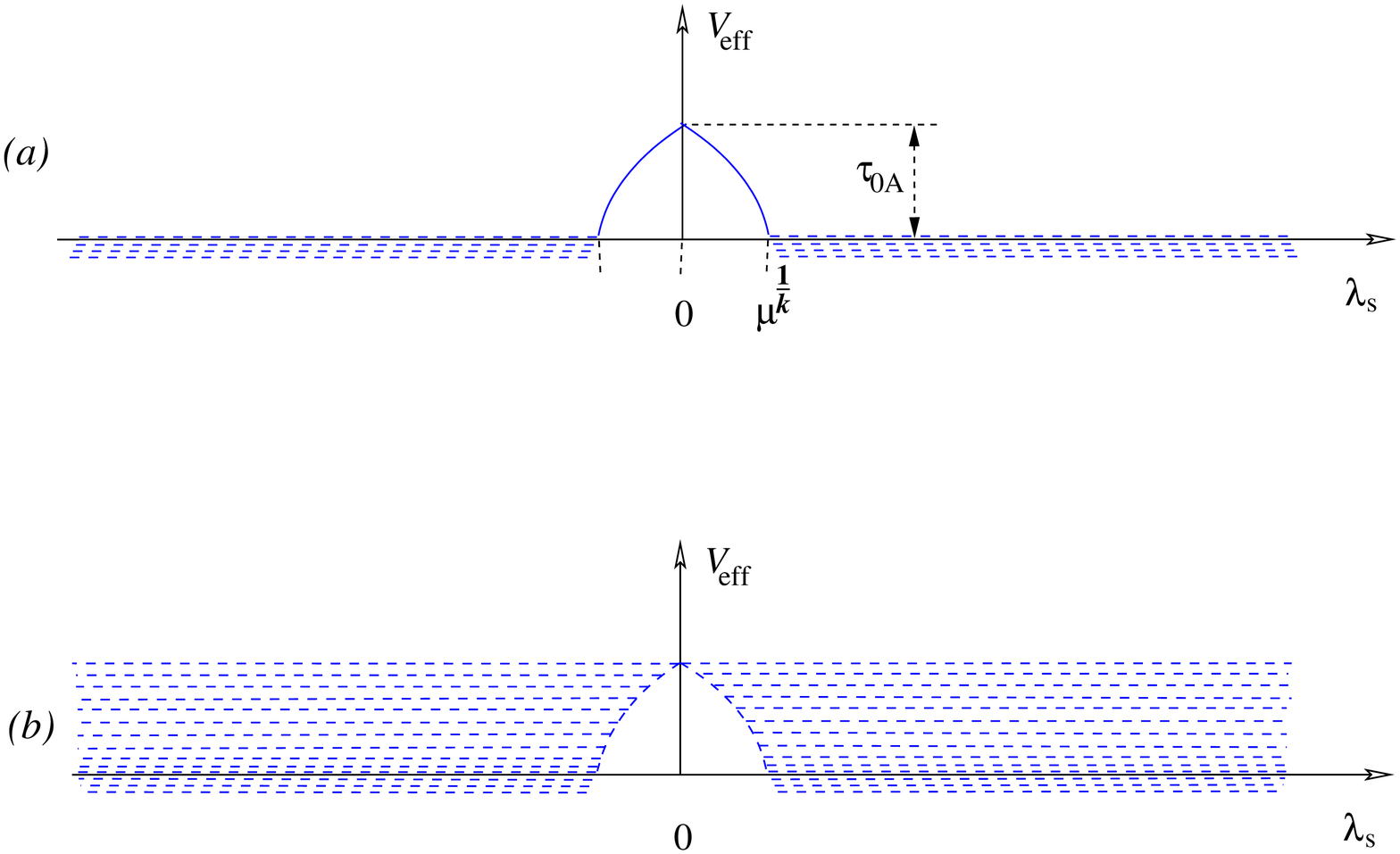}
\end{center}
\caption{\small {\it (a)} Sketch of the effective potential on the covering space for arbitrary $k$. {\it (b)} As $\mu$ decreases, the eigenvalues fill up closer to the top of the potential, decreasing the tension of the available D--brane, until at $\mu=0$, there is no brane available. }
\label{fig:doubledpot}
\end{figure}

The equation~\reef{eq:nonpert} also arises from matrix models in a
very natural manner. It first appeared in the context of complex
matrix models\cite{Dalley:1992qg}, but can be derived from Hermitian
matrix models as well. The key point is that the complex matrix models
can be formulated as usual in terms of eigenvalues, $\lambda$, of the
matrix $M^\dagger M$, which are positive.  Therefore the problem
naturally places a boundary or ``wall'' at $\lambda=0$.  The large
positive $z$ regime corresponds to the distribution pulling way from
the wall, and so the system resembles the eigenvalue distributions of
the one--cut Hermitian matrix models and yields the same perturbation
theory. The large negative $z$ limit is very different however, as the
distribution pushes up against the wall, and gives very different
physics which we will discuss at length shortly.  Another way to look
at it is in the space ${\tilde\lambda}$ where
${\tilde\lambda}^2=\lambda$. This is the natural eigenvalue space of
one matrix. Then we see that there are two identical eigenvalue
distributions, one on ${\tilde\lambda}\in[0,+\infty)$ and the other (a
mirror image) on ${\tilde\lambda}\in(-\infty,0]$. See
figure~\ref{fig:doubledpot}{\it (a)}. These distributions either pull
apart ($z\to+\infty$) or push together $z\to-\infty$.
For completeness, it is worth recalling how to derive the string
equation. We choose the case of an Hermitian matrix model, and we
simply introduce a boundary in the eigenvalue space. One can define
polynomials orthogonal to the measure $d\mu(\lambda)=d\lambda
e^{-N/\gamma V(\lambda)}$ on the interval $[-\Sigma,\Sigma]$,
modifying equation~\reef{eq:polys}:
\begin{equation}
\int_{-\Sigma}^\Sigma 
P_n(\lambda)P_m(\lambda) d\mu(\lambda)=h_n\delta_{nm}\ .
  \labell{eq:polystwo}
\end{equation}
The non--trivial (for even potential $V(\lambda)$) relations
generalizing equation~\reef{eq:master} (or more specifically, the form
given in equation~\reef{eq:masteralt})
are\cite{Dalley:1991xx}:
\begin{eqnarray}
<n-1|V^{'}({\hat \lambda})|n>&=&\frac{\gamma n}{N\sqrt{R_n} }+\frac{2\gamma}{N}\frac{P_n(\Sigma)P_{n-1}(\Sigma)}{\sqrt{h_nh_{n-1}}}e^{-\frac{N}{\gamma}V(\Sigma)}\ ,\nonumber\\
<n|{\hat \lambda}V^{'}({\hat \lambda})|n>&=&
\frac{\gamma}{N}\left((2n+1)-2\Sigma\frac{P_n(\Sigma)^2}{h_n}e^{-\frac{N}{\gamma}V(\Sigma)}\right)\ ,
  \labell{eq:mastertwo}
\end{eqnarray}
where we have used that, for even potentials,
$P_n(-\lambda)=(-1)^nP_n(\lambda)$. We can eliminate the  boundary terms to give:
\begin{eqnarray}
\frac{1}{4}\left(\frac{\gamma(2n+1)}{N}-<n|{\hat \lambda}V^{'}({\hat \lambda})|n>\right)\left(\frac{\gamma(2n-1)}{N}-<n-1|{\hat \lambda}V^{'}({\hat \lambda})|n-1>\right)\nonumber\\
=\Sigma\left(\frac{\gamma n}{N\sqrt{R_n} }-<n-1|V^{'}({\hat \lambda})|n>\right)^2\ .
\labell{eq:masterthree}
\end{eqnarray}
Note that
\begin{equation}
 <n|{\hat \lambda}V^{'}({\hat \lambda})|n>=\sqrt{R_n}<n|{\cal A}^\dagger V^{'}({\hat \lambda})|n>+\sqrt{R_{n+1}}<n|{\cal A}V^{'}({\hat \lambda})|n>\ , 
\end{equation}
and move to the continuum variables as before, defining
\begin{equation}
  \Omega(x)=\gamma x -<x-\epsilon|V^{\prime}({\hat\phi})|x>\ .
\end{equation}
Then the equation becomes:
\begin{equation}
  \labell{eq:bigstring}
\left[\Omega(x)+\Omega(x+\epsilon)\right]\left[\Omega(x-\epsilon)+\Omega(x)\right]=4\frac{\Sigma}{R(x)}\Omega^2(x)\ .
\end{equation}
Using the scalings given in equation~\reef{eq:critical}, the critical
potentials given in equation~\reef{eq:criticalpots}, and placing the
walls at the ends of the critical eigenvalue density at $\Sigma=R_c$
we get the string equations~\reef{eq:nonpert} in the limit
$\delta\to0$, having appeared at first non--vanishing order
$\delta^{4k+2}$.

\section{Matrix Models and Double Scaling III: Including Open Strings}
\label{sec:matrixopen}
In this section, we discuss how the full string
equations~\reef{eq:nonpertbos} and~\reef{eq:nonpert}, with all
parameters switched on, arise in the matrix models.

\begin{figure}[ht]
\begin{center}
    \includegraphics[scale=0.55]{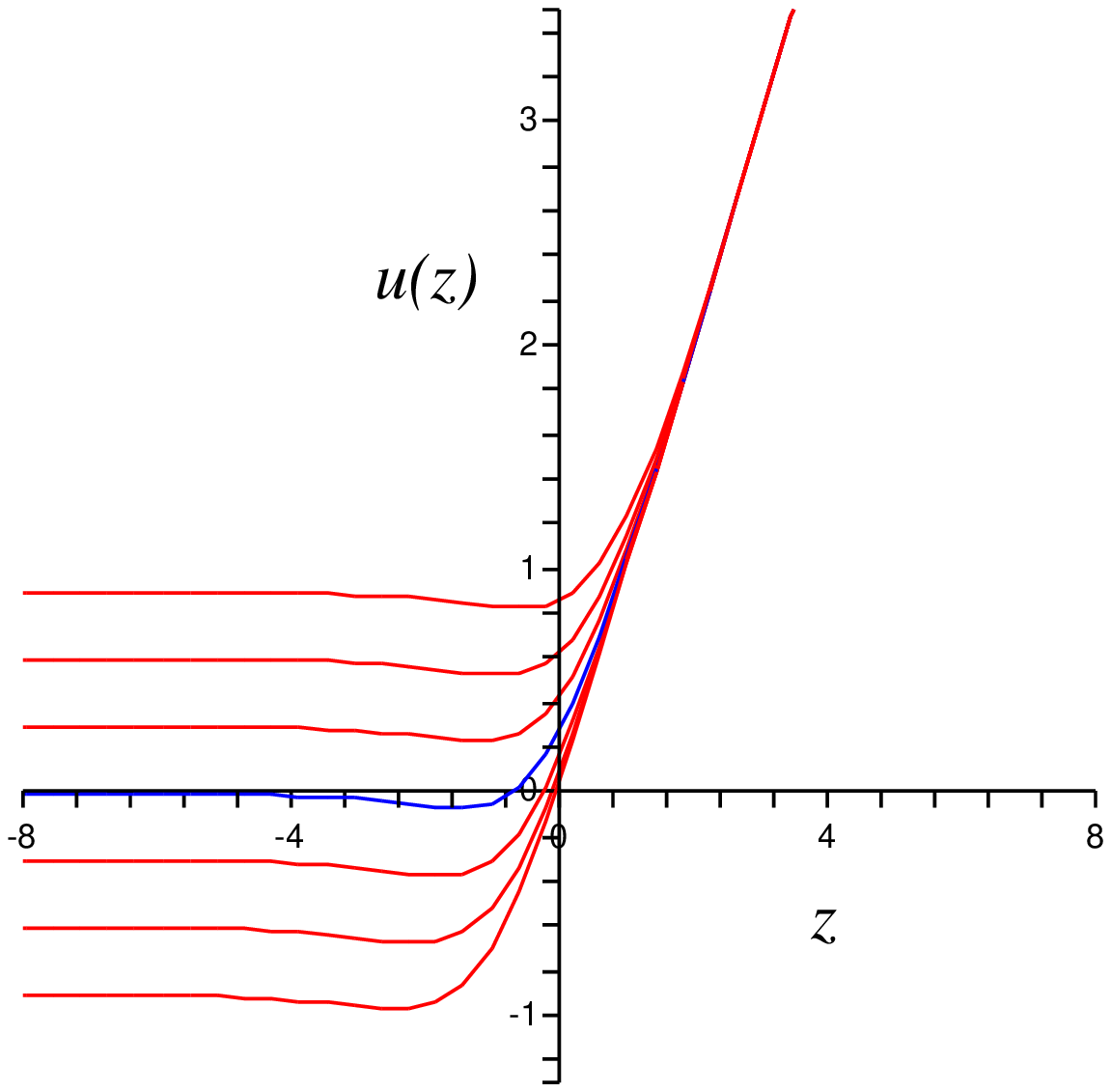}\includegraphics[scale=0.55]{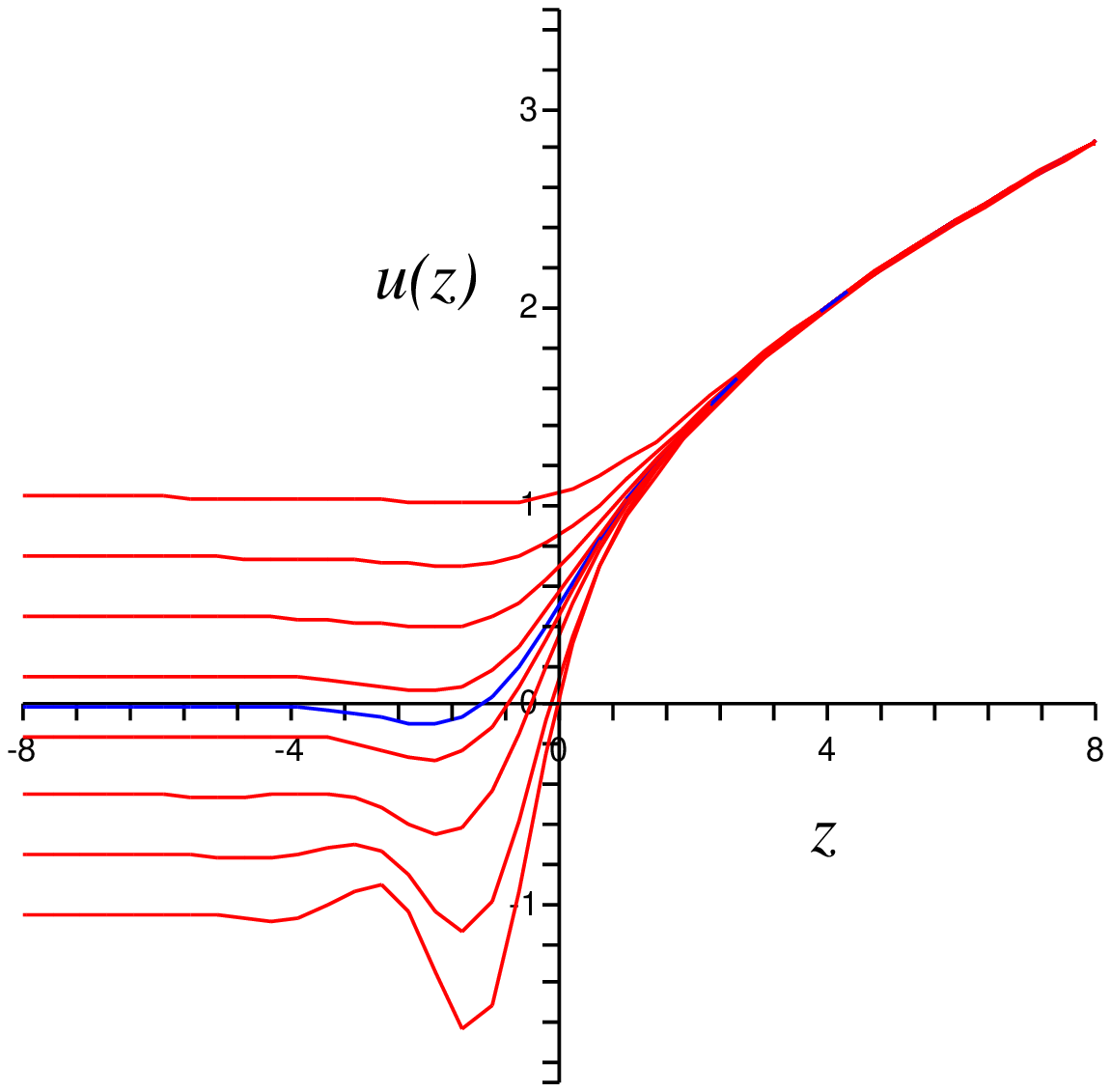}
\end{center}
\caption{\small Solutions to equation~\reef{eq:nonpert} for $u(z)$, with asymptitics given in figure~\ref{fig:gammaplots}. {\it (a)} The case $k=1$, and {\it (b)} the case $k=2$, for   a range of values of the parameter $\sigma$ with $\Gamma=0$. The middle curve in each case is the case of $\sigma=0$. }
\label{fig:sigmaplots}
\end{figure}

Let's first turn to non--zero $\sigma$ appearing in
equation~\reef{eq:nonpert}. In fact, $\sigma$ naturally appears in the
problem as the scaled position of the wall, which should be a physical
parameter as well\cite{Dalley:1991xx}. In taking the double scaling
limit, scaling the position according to
$\Sigma=R_c(1-\sigma\delta^2/2)$, yields the
equation~\reef{eq:nonpert} with non--zero $\sigma$. Well behaved
solutions for non--zero sigma are known to exist, and their behaviour
is given numerically in figure~\ref{fig:sigmaplots}.
Next, let us note that there is an important quantity which arises in
the continuum limit of the matrix models which will play an important
role here and later.  In the bra--ket formalism introduced earlier,
the operator ${\hat \lambda}$ in equation~\reef{eq:operatorchange} can
be directly formulated as a differential operator\cite{Gross:1990aw}.
This comes because the shift operators ${\cal A}, {\cal A}^\dagger$
get represented as derivative operators
$\exp\left(\epsilon\partial_x\right),
\exp\left(-\epsilon\partial_x\right)$ acting on functions of $x=n/N$.
Recall also that there is a specific scaling for the objects from
which ${\hat \lambda}$ is made, given in equation~\reef{eq:critical},
 so we can expand it and see what its scaling is in the limit
$\delta\to0$ and we find that:
\begin{equation}
  \labell{eq:criticallambda}
  {\hat\lambda}=2\sqrt{R_c}+\frac{1}{\sqrt{2}}\left(\nu^2\frac{\partial^2}{\partial z^2}-u(z)
\right)\delta^2+\cdots={\hat \lambda}_c-\frac{\cal H}{\sqrt{2}}\delta^2+\cdots
\end{equation}
The Schr\"odinger operator ${\cal H}$ plays a crucial and very natural
role in what is to come.

The matrix model origin of the string equation~\reef{eq:nonpertbos}
with non--zero $\Gamma$ arises by adding a
term\cite{Kazakov:1990cq,Kostov:1990nf}
\begin{equation}
  \labell{eq:extrabit}
  \Delta V(M)=\frac{\Gamma}{N}\Tr\log \left(1-s^2M^2\right)
\end{equation}
to the matrix model potential. In the world--sheet dual to the Feynmann
diagrams, this corresponds to adding holes of all sizes, by expansion
of the logarithm. This new term adds new pieces to the master
equation~\reef{eq:masteralt}:
\begin{equation}
  \labell{eq:newterms}
  \frac{\Gamma}{N}\left(1-<n|\frac{1}{1-s^2{\hat\lambda}^2}|n>\right)\ .
\end{equation}
On the sphere, there is still the critical behaviour where large area
surfaces dominate, near $R=R_c$, but there is an additional critical
point near a critical value of $s$ set by $s_c^{-1}=2\sqrt{R_c}$. This
arises because, {\it e.g.,} on the sphere, the last term in
equation~\reef{eq:newterms} reduces to
$-(\Gamma/N)(1-4s^2R(x))^{-1/2}$.  So the lengths of loops can chosen
to diverge if we send $s\to s_c$, but give finite length loops as
$\delta\to0$ if we scale at the right rate. Since near the critical
point the new terms are:
\begin{equation}
  \label{eq:nearcritical}
  -\frac{\Gamma}{N}\frac{1}{[4(s^2_c-s^2)R_c+(R_c-R)/R_c]^{\frac{1}{2}}}\ ,
\end{equation}
this rate turns out to be $s^2=s^2_c-\sigma\delta^2$. The factor of
$1/N$ scaling as $\delta^{2k+1}$ produces the appropriate damping of the
divergence and the string equation~\reef{eq:nonpertbos} results (as
before) at order $\delta^{2k}$, the resolvent arising from the appearance
of ${\cal H}$ in the scaling part of ${\hat\lambda}$, appearing at the
same order in $\delta$ as $\sigma$ does.

The same procedure can be applied to the master
equations~\reef{eq:masterthree} arising from placing walls in the
system. As happens before, order $\delta^{4k+2}$ is where the
non--trivial physics appears, and the lefthand side of the string
equation~\reef{eq:nonpert} appears at this order, while the right hand
side, $\nu^2\Gamma^2$ appears multiplying $-1$ times the left hand
side of the Gel'fand--Dikii equation~\reef{eq:resolventequation}, with
$\lambda$ set to $\sigma$. Thus we arrive at
equation~\reef{eq:nonpert}. So this is how the position of the walls
and the weight factor of the loops become related in the continuum
model. In effect, the argument of the logarithm diverges (so as to
give finite length loops in the continuum limit) at the same place in
eigenvalue space at which the walls are located.

Another way to see that this must be true is to imagine that
equation~\reef{eq:nonpert} was true, but for another parameter $\tau$,
instead of $\sigma$.  How can we fix $\tau$? Taking sums and
differences of the equations we see that we have a condition that
either $\tau=\sigma$ and the equation~\reef{eq:nonpert} follows, or we
have ${\cal R}=0$, in which case $\Gamma=0$. This matches what we
learned from the matrix model.  If we do not scale the parameters so
as to make the logarithm diverge precisely at the ends, then it
produces no new physics.

\section{Effective Tachyon Potentials and D--Branes in Type 0A String Theory}  
 \label{sec:potentialsagain} 
 
 Turning to the issue of instantons arising from analyzing
 perturbation expansions, we note\cite{Dalley:1992vr} that there is
 another instanton contribution to the large positive $z$ perturbation
 theory. Searching as before for the perturbation $\epsilon=e^{-f}$,
 we get the same instanton as for the equation~\reef{eq:nonpertbos},
 and in addition, there is an equation:
  \begin{equation}
    \labell{eq:newinstanton}
    (u_0-\frac{\nu^2}{4}(f^{'})^2){\cal R}^2=0\ ,
  \end{equation}
which, upon using $u_0=z^{\frac{1}{k}}$, gives instanton:
\begin{equation}
  \labell{eq:instant}
  \epsilon\sim e^{-\frac{4k}{2k+1}\frac{1}{g_s}}\ .
\end{equation}
This instanton can be connected to the physics of the effective
potentials we have already studied.  See figures~\ref{fig:effpots}
and~\ref{fig:doubledpot}{\it (a)}. For large positive $z$, the results
from before will still apply (after a shift of the origin of
coordinates to place one of the walls there), modified only by the
fact that the eigenvalues must be positive. Looking at the effective
potentials, we see that since there is a wall at $\lambda_s=0$, the
instanton at negative $\lambda_s$ previously encountered for $k$ even
cannot destabilize the system.  There is an instanton for all $k$ at
the wall, and its tension is given by $V^{(k)}_{\rm eff}(0)$, which a
computation shows is:
\begin{equation}
  \labell{eq:tensionzeroay}
  \tau_{0A}=V_{\rm eff}(\lambda_s=0)=\frac{4k}{2k+1}\frac{\mu^{1+\frac{1}{2k}}}{\nu}\ .
\end{equation}
This is the same action as the new instanton perturbations of the
solutions we saw above. So we see that although the same instanton
which destabilised the bosonic theory is present in the analysis, it
does not represent an instability\cite{Dalley:1992vr}.

Since the role of $\sigma$ in the matrix model is to move the wall, as
we saw in the previous section, it can be seen as an interpolating
parameter in some sense, since  the limit of large negative $\sigma$
approaches (but never reaches) the physics of the unstable bosonic
models\cite{Johnson:1992wr}. See figure~\ref{fig:sigmaplots}{\it (b)}.
Notice for example how as $\sigma$ becomes more negative, for the case
$k=2$, the system begins to develop more fluctuations at negative $z$,
ultimately heralding the poles that exist on the negative $z$ axis for
the Painlev\'e~I equation (equation~\reef{eq:nonpertbos} for $k=2$.).
Studies of this sort were presented in ref.\cite{Johnson:1992wr}.

The physical role of $\sigma$ from the point of the continuum string
theory is more subtle. It represents a non--perturbative contribution
to the coupling of a boundary operator ${\cal O}_{\rm B}$ which
measures the length of world--sheet boundaries or
loops\cite{Johnson:1992wr}. It is a boundary cosmological constant and
we will examine it in some more detail in the next section. Note that
for the models defined by equation~\reef{eq:nonpertbos} with
$\Gamma=0$, the boundary length operator for the $k$th model is
identified with ${\cal O}_{k-1}$. See ref.\cite{Martinec:1991ht}. In
particular, for $k=1$, the operator is ${\cal O}_0$ which couples to
$t_0\sim z$. For this case, (and for the $k=1$ solution of
equation~\reef{eq:nonpert} for large positive $z$) it is interesting
to note that $\sigma$ can be absorbed into $z$ without changing the
universal physics. So we see that $\sigma$ is indeed a natural
realization of the boundary cosmological constant, by matching onto
the perturbatively identified boundary operator.

Recall that the $k=1$ model has no expansion in terms of closed string
world--sheets. The theory's entire content comes alive when the other
operators ${\cal O}_k$ are inserted, and their correlation functions
at genus $h$ are then seen to be isomorphic to intersection numbers on
the moduli space of Riemann surfaces of genus $h$: it is a topological
theory. The $k=1$ solution of equation~\reef{eq:nonpert} (for large
$z$) shares this behavior, but develops new features in the
non--perturbative regime. In fact, this is a precise example of a
non--perturbative completion of a topological theory, and deserves
further study from this perspective. It can be rewritten as the
type~0B model in the same background, as identified in
ref.\cite{Klebanov:2003wg} using the change of variables of
refs.\cite{Morris:1990bw,Morris:1992zr}.

As we will recall below the partition function of the $k=1$ theory
(exactly for equation~\reef{eq:nonpertbos} and at large positive $z$
for equation~\reef{eq:nonpert}) only has a topological expansion in
the presence of D--branes, giving only world--sheets with boundaries
in the expansion.

\section{Loops and D--Branes for Bosonic and Type 0A Strings}
\label{sec:loopsbranes}
Now we turn to the physics of non--zero $\Gamma$. A study of
equation~\reef{eq:nonpert} for non--zero $\Gamma$ shows that large
positive $z$ (or $\mu$) perturbation theory has an interpretation as
including world--sheet boundaries, and in fact $\Gamma$ counts the
number of D--branes in the background, as can be seen by the fact that
a factor $\Gamma^b$ appears for a surface with the number of
boundaries equal to $b$. For example, for $k=2$:
\begin{eqnarray}
F&=&\frac{4}{15}g_s^{-2}+\frac{8\Gamma}{5} g_s^{-1}+\frac{(6\Gamma^2+1)}{24}\log \mu - \frac{(4\Gamma^3+3\Gamma)}{48} g_s-\frac{1}{5760}(180\Gamma^4+345\Gamma^2+28)g_s^2+\cdots\nonumber\\
&=&\sum_{h,b}C_{b,h}g_s^{2h-2+b}\Gamma^b\ .
  \labell{eq:boundaries}
 \end{eqnarray}
 The equation supplies fully non--perturbative completions of this
 physics, including the following behaviour at large negative $z$ (or $\mu$):
\begin{eqnarray}
F&=&-\frac{(4\Gamma^2-1)}{4}\log\mu \labell{eq:fluxes}
 \\
&&+(4\Gamma^2-1)(4\Gamma^2-9)(4\Gamma^2-25)\left\{\frac{1}{960}g_s^4+\frac{1}{30720}(48\Gamma^4-1240\Gamma^2+8371)g_s^8\right\}+\cdots
\ ,\nonumber
 \end{eqnarray}
 where the first term is universal to all $k$. At face value, the
 interpretation in terms of boundaries can be made in this regime as
 well, but explanation for the number of exactly vanishing orders in
 perturbation theory (all surfaces with an odd number of boundaries,
 for example) is needed. Instead, ref.\cite{Klebanov:2003wg} pointed
 out that in this regime, there are no branes, but instead R--R
 fluxes, using facts from the continuum type 0 theories and the matrix
 model.  Here, we'll study the matrix models further and uncover the
 effective potential beyond tree level in both regimes. Its behaviour
 in the different regimes clearly give support to this interpretation.

 The large positive $z$ perturbation theory of the equation matches
 perturbation theory obtained from equation~\reef{eq:nonpertbos},
 which for non--zero $\Gamma$ and $\sigma$ was formulated in
 refs.\cite{Kazakov:1990cq,Kostov:1990nf}.  $\Gamma$ appears
 multiplying the resolvent, which in turn satisfies
 equation~\reef{eq:resolventequation}. In fact, the most efficient way
 to construct a perturbative expansion for $u(z)$ is to treat
 equation~\reef{eq:nonpertbos} as an equation for ${\hat R}$ and then
 substitute it into the resolvent equation, yielding an equation for
 ${\cal R}$ and hence for $u(z)$. This of course yields the
 equation~\reef{eq:nonpert}. Of course, this larger equation allows
 solutions for $u(z)$ which are not solutions of
 equation~\reef{eq:nonpertbos}. These are the solutions which have the
 well defined large negative $z$ asymptotics and accompanying pleasant
 non--perturbative behaviour which we now associate to type 0A. See
 the families of curves displayed in figure~\ref{fig:gammaplots}.
 
 That the resolvent appears explicitly in the description of D--branes
 in the backgrounds is interesting because the same object is known to
 control the behaviour of macroscopic loops in the continuum
 theory\cite{Banks:1990df}. This is worth recalling in some detail: In
 the matrix model, it is natural to consider the Wilson loop:
\begin{equation}
  \labell{eq:Wilson}
 W(\ell)=\langle\frac{1}{N}\Tr e^{\ell M}\rangle\ .
\end{equation}
Its Laplace transform is
\begin{equation}
  \labell{eq:resolve}
  W(\lambda)=\langle\frac{1}{N}\Tr\frac{1}{\lambda-M}\rangle=\int_0^\infty d\ell e^{-P\ell}W(\ell)\ .
\end{equation}
and in the  continuum limit it is:
\begin{equation}
  \labell{eq:continuumloop}
  W(\lambda)=\int d\xi\frac{\rho(\xi)}{\lambda-\xi}\ .
\end{equation}
The saddle point equation of the matrix model is:
\begin{equation}
  \labell{eq:motion}
  \frac{V^\prime(\lambda)}{2}={\bf P}\int_{\cal C}d\xi\frac{\rho(\xi)}{\lambda-\xi}\ ,
\end{equation}
where $\bf P$ denotes the principal part and the integral is over the
support ${\cal C}$ of the density function, $\rho(\lambda)$, of
eigenvalues $\lambda$.  The density is normalized according to $\int
d\lambda \rho(\lambda)=1$.  The real part of $W$ is equal to
$V^{\prime}/2$ on the support of $\rho(\lambda)$ where its imaginary
part is non--zero.  This is nicely expressed by the equation (see
ref.\cite{Makeenko:2002uj} for a useful presentation):
\begin{equation}
  \labell{eq:equationquad}
 {\rm Im}[P(\lambda)]= {\rm Im}(V^{\prime}W-W^2)=(V^{\prime}-2{\rm Re} W){\rm Im} W=0
\end{equation}
with solution
\begin{equation}
  \labell{eq:equationloop}
  W(\lambda)=\frac{V^\prime(\lambda)}{2}\pm\frac{1}{2}\sqrt{(V^\prime)^2-4P(\lambda)}\ .
\end{equation}
where $P(\lambda)$ is fixed by the asymptotic $W(\lambda)\to-1/\lambda$ as
$\lambda\to\infty$, given the normalization of the density. By
definition, $\rho(\lambda)$ is given by the discontinuity of
$W(\lambda)$ on the cut ${\cal C}$.

The connection with the continuum loop operator and the continuum
resolvent operator comes from the definition:
\begin{equation}
  \labell{eq:continuumdiscrete}
  W(\lambda)=\frac{V^\prime}{2}+\delta^{ 2k-1}w(\lambda_s)
\end{equation}
where, as before, $\lambda_s$ is the scaling part of $\lambda$.

Now the marvellous thing\cite{Banks:1990df} is that 
\begin{equation}
  \labell{eq:expect}
  w(\lambda_s)=\int\! dz\, {\hat R}(z,\lambda_s)=\int dz<z|\frac{1}{({\cal H}-\lambda_s)}|z>\ ,
\end{equation}
and it is the Laplace transform of the continuum loop
\begin{equation}
  \labell{eq:cont}
  w(\ell)=\int dz <z|e^{-\ell {\cal H}} |z>\ ;\qquad w(\lambda_s)=\int_0^\infty d \ell
e^{-\lambda_s\ell}w(\ell)
\end{equation}
The Schr\"odinger operator ${\cal H}$ arose earlier in
equation~\reef{eq:criticallambda}. Loops correspond to expectation
values of traces of powers of matrices $\Tr{M^p}$ where
$p=\frac{\ell}{\delta}$. This corresponds to cutting a $p$--sided hole
in the dual picture of discretized worldsheets. We hold~$\ell$ fixed
by sending $p$ to infinity as we send $\delta\to0$.  In the bra--ket
formalism this amounts to computing the Laplace transform of the
resolvent ${\hat R}(z,\lambda_s)$, and we integrate over $z$ (up to
some reference value, $\mu$ although we will still use $z$ for a lot
of what follows) to get the expectation value $w(\ell)\sim\int dz <z|
{\hat\lambda}^{\frac{\ell}{\delta}}|z>$, which gives the first
expression in equations~\reef{eq:cont}.

Notice that $w(\lambda_s)$ is also (up to a factor $\nu/2$) the first
$\lambda_s$--derivative of the effective potential $V_{\rm
  eff}(\lambda_s)$, and this gives an efficient way of deriving the
effective potentials for the system to arbitrary order in perturbation
theory, by use of the string equations in combination with the
resolvent equation.  In particular, the procedure is to solve either
string equation for $u(z)$ to desired order in perturbation theory,
keeping some non--zero value of $\sigma$. This is now to be treated as
a potential for the Hamiltonian ${\cal H}$, and one can solve for its
resolvent ${\hat R}(z,\lambda_s)$ to that order in perturbation theory
by using equation~\reef{eq:resolventequation}. The next step is to
integrate once with respect to $z$ to get the loop operator
$w(\lambda_s)$, and one can integrate once with respect to $\lambda_s$
to get the effective potential $V_{\rm eff}$:
\begin{equation}
  \labell{eq:effectivepotentialint}
V_{\rm eff}=\frac{2}{\nu}\int w(\lambda_s)d\lambda_s \ .
\end{equation}
In fact, it is more interesting to examine the
properties of $w(\lambda_s)$, for its zeros tell us about the possible
location of D--branes. We already know the structure of the zeros at
the level of the sphere in perturbation theory, where $\Gamma$ is
invisible. This procedure allows us to efficiently examine the effect
of non--zero $\Gamma$ on the structure of the zeros. It also allows us
to study the large negative $z$ behaviour of the effective potential
rather clearly. The previous methods were limited to the sphere,
and in particular the leading behaviour on the sphere vanishes for
large negative $z$.

To start with, take the case $k=1$. For large positive $z$, we have,
to disc level:
\begin{equation}
  \labell{eq:kone}
  u(z)=z+\frac{\nu\Gamma}{\sqrt{z-\sigma}}+\cdots
\end{equation}
This can be substituted into equation~\reef{eq:resolventequation}, and
solve for the resolvent to the same order in perturbation theory. The
result is:
\begin{equation}
  \labell{eq:koneresolve}
  {\hat R}(z,\lambda_s)=\frac{1}{2}\frac{1}{(z-\lambda_s)^{\frac{1}{2}}}-\frac{1}{4}\frac{\nu\Gamma}{(z-\lambda_s)^{\frac{3}{2}}(z-\sigma)^{\frac{1}{2}}}+\cdots\ ,
\end{equation}
and this results in, after integrating with respect to $z$:
\begin{equation}
  \labell{eq:koneloop}
  w(\lambda_s)=(\mu-\lambda_s)^{\frac{1}{2}}+\frac{1}{2}\frac{\nu\Gamma (\mu-\sigma)^{\frac{1}{2}}}{(\lambda_s-\sigma)(\mu-\lambda_s)^{\frac{1}{2}}}+\cdots\ ,
\end{equation}
and an integration with respect to $\lambda_s$:
\begin{equation}
  \labell{eq:koneveff}
  V_{\rm eff}(\lambda_s)=\frac{4}{3\nu}(\mu-\lambda_s)^{\frac{3}{2}}+2\Gamma\tanh^{-1}\left( \frac{\mu-\lambda_s}{\mu-\sigma}\right)^{\frac{1}{2}}\ .
\end{equation}

Look at the zeros of $w(\lambda)$, we find that we must solve the equation:
\begin{equation}
  \labell{eq:konezeros}
  (\lambda_s-\sigma)(\mu-\lambda_s)+\frac{1}{2}\nu\Gamma(\mu-\sigma)^\frac{1}{2}=0\ .
\end{equation}
This is a cute equation.  At the level of the sphere, ($\Gamma=0$ or
$\nu=0$), it is a quadratic, and we see that we have the roots
$\lambda_s=\sigma,\mu$. This is what we had before, recalling that
$\sigma$ marks the position of the wall, and $\mu$ is where the
eigenvalue density $\rho(\lambda_s)$ ends. Note that the value of the
effective potential at $\mu$ is zero, while at $\lambda_s=\sigma$ it is
\begin{equation}
  \labell{eq:tension}
  V_{\rm eff}(\sigma) =\frac{4}{3}\frac{(\mu-\sigma)^{\frac{3}{2}}}{\nu}\ .
\end{equation}
Note that at $\sigma=0$, this gives $\tau_{\rm 0A}$.  Beyond the
sphere we see that $\Gamma$ deforms the quadratic, changing the
location of the zeros:
\begin{equation}
  \labell{eq:konelambda}
  \lambda_s=\frac{\sigma+\mu}{2}\pm\frac{1}{2}\sqrt{(\sigma-\mu)^2+2\nu\Gamma\sqrt{\mu-\sigma}}\ .
\end{equation}

To get a feeling for the effects of the parameter $\Gamma$, let us
work at large $\mu$, in which case $w(\lambda_s)$ becomes:
\begin{equation}
  \labell{eq:largezw}
  w(\lambda_s)=\mu^{\frac{1}{2}}-\frac{1}{2}\frac{\lambda_s}{\mu^{\frac{1}{2}}}+\frac{1}{2}\frac{\nu\Gamma}{\lambda_s-\sigma}+\frac{1}{4}\frac{\nu\Gamma}{\mu}\ ,
\end{equation}
with zeros at approximately:
\begin{equation}
  \labell{eq:zeroslargezw}
  \lambda_s^-=\sigma-\frac{1}{2}\frac{\nu\Gamma}{\mu^{\frac{1}{2}}}\ ,\quad \lambda_s^+=\mu+\frac{1}{2}\frac{\nu\Gamma}{\mu^{\frac{1}{2}}}\ .
\end{equation}

The zero to the right of $\mu$ lies on a cut which runs to infinity,
marking support of the eigenvalue distribution. The other zero can be
thought of as a cut which has collapsed to zero size, in the language
of ref.\cite{Klebanov:2003wg}.  As explained in
ref.\cite{Klebanov:2003wg}, there is a collapsed Riemann surface of
the form $y^2=f(\lambda_s)$ where $y=w(\lambda_s)$, extending to the
complex $\lambda_s$--plane.  The vanishing of $w(\lambda_s)$ controls
the location of the cuts on the complex plane $\lambda_s$, and the
surface is the double cover of the plane joined through the cuts,
giving a surface of specific topology. We can perform the integral
$(1/\pi i \nu)\oint w(\lambda_s) d\lambda_s$ along a closed contour
running between the cuts on one sheet and closing the loop by going
along the other sheet. The normalization is chosen to compute the
change in the effective potential $V_{\rm eff}$, following
equation~\reef{eq:effectivepotential}. For our problem in hand, we see
that the result comes from the pole at $\lambda_s=\sigma$, and the
result is precisely $\Gamma$. This counts the number of branes in the
problem.

We've computed a pleasant general structure for the higher $k$ cases.
The leading contribution (up to disc order) to $u(z)$ is:
\begin{equation}
  u(z)=z^{\frac{1}{k}}+\frac{1}{k}\frac{\nu\Gamma}{z^{\frac{k-1}{k}}(z^{\frac{1}{k}}-\sigma)^{\frac{1}{2}}}+\cdots\ ,
\end{equation}
and some algebra produces the resolvent:
\begin{equation}
  \labell{eq:resolved}
  {\hat R}(z,\lambda)=\frac{1}{2}\frac{1}{(z^{\frac{1}{k}}-\lambda_s)^{\frac{1}{2}}}-\frac{1}{4k}\frac{\nu\Gamma}{(z^{\frac{1}{k}}-\lambda_s)^{\frac{3}{2}} (z^{\frac{1}{k}}-\sigma)^{\frac{1}{2}}z^{\frac{k-1}{k}}}+\cdots
\end{equation}
Integrating this up gives
\begin{eqnarray}
w^{(k)}(\lambda_s)&=&k!\sum_{n=1}^k\frac{2^{n-1}}{(k-n)!(2n-1)!!}(\mu^\frac{1}{k}-\lambda_s)^{n-\frac{1}{2}}\mu^{1-\frac{n}{k}}
+\frac{1}{2}\frac{\nu\Gamma(z^{\frac{1}{k}}-\sigma)^{\frac{1}{2}}}{(\lambda_s-\sigma)(z^\frac{1}{k}-\lambda_s)^{\frac{1}{2}}}+\cdots\
,
 \labell{eq:generalkresultsw}
\end{eqnarray}
 and
\begin{eqnarray}
V_{\rm
  eff}^{(k)}(\lambda)&=& \frac{1}{\nu}\sum_{n=1}^k\frac{k!(-2)^{n+1}}{(k-n)!(2n+1)!!}(\mu^{\frac{1}{k}}-\lambda_s)^{n+\frac{1}{2}}\mu^{1-\frac{n}{k}}+2\Gamma\tanh^{-1}\left(\frac{\mu^{\frac{1}{k}}-\lambda_s}{\mu^{\frac{1}{k}}-\sigma}\right)^{\frac{1}{2}}+\cdots\ ,
  \labell{eq:generalkresultsveff}
\end{eqnarray}
where we reproduce the previous effective potential expressions on the
sphere, but have in addition a disc level contribution.

The location of the zeros of $w^{(k)}(\lambda_s)$ give the extrema of
the effective potential. The issue boils down to a $(k+1)$th order
polynomial in $\lambda_s$, which on the sphere has zeros at $\sigma$
and $\mu^{\frac{1}{k}}$ for positive $\lambda_s$, and for odd $k$ there
are no other real ones (only $k-1$ imaginary ones). For even $k$ there
are $k-1$ imaginary ones and a single extra one at $-c
\mu^{\frac{1}{k}}$.  At the level of the disc these zeros get deformed,
as before, but notice that  there is again a
pole at $\lambda=\sigma$, which will give a non--vanishing
contribution of precisely $\Gamma$ to the B--cycle integral $(1/\pi i \nu)
\oint w(\lambda_s)d\lambda_s$ between the cut to the right of
$\lambda_s=\mu^{\frac{1}{k}}$ and the collapsed cut located just to the left of
$\lambda_s=\sigma$.

There are presumably means of deforming the problem in order to have
non--zero B--cycle integrals between the cut and the other $(k-1)$
zeros, and these would define other branes too. It is not clear
whether this can be done in this matrix model description.

We can just as easily study and construct the same quantities in the
negative $z$ limit too.  In fact, we can write everything succinctly
for all $k$ to the first non--vanishing order. We have in the large
negative $z$ direction:
\begin{equation}
u(z)=\sigma+\frac{\nu^2}{4}\frac{(4\Gamma^2-1)}{(z-\sigma^k)^2}+\cdots
  \labell{eq:largezkone}
\end{equation}
Putting this into the resolvent equation~\reef{eq:resolventequation},
we get
\begin{equation}
  \labell{eq:negativeresolve}
  {\hat R}(z,\lambda_s)=\frac{1}{2}\frac{1}{(\sigma-\lambda_s)^{\frac{1}{2}}}-\frac{1}{16}\frac{\nu^2(4\Gamma^2-1)}{(\sigma-\lambda)^{\frac{3}{2}}(z-\sigma^k)^2}+\cdots
\end{equation}
From this we get:
\begin{equation}
  \labell{eq:wnegative}
w(\lambda_s)=\frac{1}{2}\frac{\mu}{(\sigma-\lambda)^{\frac{1}{2}}}+\frac{1}{16}\frac{\nu^2(4\Gamma^2-1)}{(\sigma-\lambda)^{\frac{3}{2}}(\mu-\sigma^k)}\ ,
\end{equation}
and 
\begin{equation}
  \labell{eq:veffnegative}
  V_{\rm eff}(\lambda_s)=\mu(\sigma-\lambda_s)^{\frac{1}{2}}-\frac{1}{8}\frac{\nu^2(4\Gamma^2-1)}{(\mu-\sigma^k)(\sigma-\lambda_s)^{\frac{1}{2}}}\ .
\end{equation}

The most interesting thing here is that the zeros are at:
\begin{equation}
  \labell{eq:zerosnegative}
  \lambda_s=\sigma-\frac{\nu^2(4\Gamma^2-1)}{8\mu(\sigma^k-\mu)}\ .
\end{equation}
There is only one zero.  On the sphere, it is at $\sigma$. The
effective potential vanishes there. So there is no sign of anything
that we would interpret as a sector representing D--branes of finite
tension. This is consistent with the interpretation that there are no
D--branes in this limit. All we have is the eigenvalue sea. In a sense
it has filled up beyond the top of the potential and so there are no
branes. This is what is depicted (somewhat 
schematically) in figure~\ref{fig:doubledpot}{\it (b)}.

Beyond the sphere does not change this conclusion. There is a single
zero, and the value of the potential does not have a $g_s^{-1}$
scaling normally associated to D--branes. It is important to note that
this is because of how equation~\reef{eq:nonpert} and the resolvent
equation works perturbatively.  First note that the resolvent
equation~\reef{eq:resolventequation} cannot itself generate odd powers
of $g_s$ or $\Gamma$ in any expansion for ${\hat R}(z,\lambda)$.  It
must get it from the potential $u(z)$ which is governed by the string
equation. It can do this from (for example) the disc term in the large
positive $z$ expansion.  Equation~\ref{eq:nonpert}, however, gives
only even powers of $g_s$ and for $\Gamma$ in a large negative $z$
expansion for $u(z)$.  This then ensures that the resolvent will never
have odd $\Gamma$ or $g_s$, and therefore the effective potential
$V_{\rm eff}(\lambda_s)$ will never give rise to energies which have
such dependence. Therefore there can be no objects which have their
mass--energy originating at disc order in perturbation theory, and
hence no D--branes for large negative $z$. This constitutes a proof of
this directly in this language, and lends further credence to
ref.\cite{Klebanov:2003wg}'s interpretation of $\Gamma$ as controlling
closed string backgrounds, namely R--R fluxes.

It is further interesting to note that if we include the next order in
perturbation theory beyond the sphere, the value of the effective
potential at the zero is:
\begin{equation}
  \labell{eq:fluxnegative}
  V^0_{\rm eff}=-\frac{(4\Gamma^2-1)^{\frac{1}{2}}|\mu|^{\frac{1}{2}}}{\sqrt{2}(\mu-\sigma^k)^{\frac{1}{2}}} +\cdots\ ,
\end{equation} 
which scales as $g_s^0$, which is the correct scaling for R--R flux
in closed string perturbation theory, or for a torus contribution. 

Finally, it is interesting to study a family of exact solutions to
equation~\reef{eq:nonpert} for each $k$. The solutions are:
\begin{equation}
u(z)=\sigma+\frac{\nu^2}{4}\frac{(4\Gamma^2-1)}{(z-\sigma^k)^2}\ ,\qquad \Gamma=m+\frac{1}{2}\ ,\quad m=0\hdots k\ .
  \labell{eq:exact}
\end{equation}
For these, the analysis above goes through as before, with similar
conclusions.  It is not known what the significance of such solutions
might be in this context, but their existence is worth remarking upon,
especially since (as noted in ref.\cite{Dalley:1992vr}) all the
solutions asymptote to them at large negative $z$, and so they may
capture some universal aspects of the physics in this regime. We make
a further observation about these solutions in the next section.

\section{Loop Equations and Virasoro Constraints}
\label{sec:loopsbranesagain}
As we've seen, the basic organizing object in the theory for much of
our considerations has been the loop operator $w(\lambda_s)$, which is
the first integral of the diagonal of the resolvent ${\hat
  R}(z,\lambda_s)$. We have studied it in the presence of D--branes
(and fluxes) in the background, which enters the matrix model string
equations in a natural way, again in terms of the resolvent ${\hat
  R}(z,\sigma)$, now in terms of the fixed background parameter
$\sigma$ and multiplied by $\Gamma$.

It is natural that the loop expectation value $w(\lambda_s)$ is
controlled by the same object that we use to place D--branes (and
fluxes, for type 0A) into the the background, and one might wonder if
this relation can be made more precise.  Can one express the open
string background as a specific combination of the existing closed
string operators?  If so, this would be a clear example of the
interchangeability of closed and open string descriptions of a given
background, perhaps the sharpest demonstration of open--closed
duality.  In fact, it can be done\cite{Itoh:1992vv,Johnson:1994vk},
and it may be illustrative of the correct language to use for other
examples.

Let us describe this using the equivalent microscopic loop operator
language. The object $w(\ell)$ can be expanded in terms of the
microscopic scaling operators
\begin{equation}
  w(\ell)=\sum_{n=0}\frac{\ell^{n+\frac{1}{2}}}{\Gamma\left(n+\frac{3}{2}\right)}{\cal
  O}_n\ ,
\end{equation}
and hence the Laplace transformed operator can be written:
\begin{equation}
  \labell{eq:loopmicro}
  w(x)=\sum_{n=0} x^{-n-\frac{3}{2}} {\cal O}_n\ .
\end{equation}
Insertion of operator ${\cal O}_n$ is equivalent to differentiation
with respect to the coupling $t_n$:
\begin{equation}
  \labell{eq:source}
  <{\cal O}_n>=\nu^2\frac{\partial F}{\partial t_n}\ .
\end{equation}
It was shown long ago\cite{Dijkgraaf:1991rs,Fukuma:1991jw} that the
purely closed string theory ($\Gamma=0,\sigma=0$) is equivalent to the
following constraints on $\tau=\sqrt{Z}=\sqrt{\exp(-F)}$:
\begin{equation}
  \labell{eq:virasoro}
  L_n\tau=0\ ,\qquad n\geq -1\ ,
\end{equation}
where
\begin{eqnarray}
L_{-1}&=&\sum_{k=1}^{\infty}\left(k+\frac{1}{2}\right)t_k\frac{\partial}{\partial
t_{k-1}}+\frac{1}{4\nu^2}z^2 \ , \nonumber\\  
L_0&=&\sum_{k=0}^{\infty}\left(k+\frac{1}{2}\right)t_k\frac{\partial}{\partial
t_k}+\frac{1}{16} \ ,\nonumber\\
L_n&=&\sum_{k=0}^{\infty}\left(k+\frac{1}{2}\right)t_k\frac{\partial}{\partial
t_{k+n}}+4\nu^2\sum_{k=1}^{n}\left(k+\frac{1}{2}\right)t_k\frac{\partial^2}{\partial
t_{k-1}\partial
t_{n-k}} \ ,
\labell{eq:constraints}
\end{eqnarray}
  which form a Virasoro algebra:
  \begin{equation}
    [L_n,L_m]=(n-m)L_{n+m}\ .
  \end{equation}
  Note that the $L_{-1}$ constraint, after taking a derivative, is
  equivalent to the string equation~\reef{eq:nonpert}:
  \begin{equation}
    \labell{eq:stringequationvir}
    -\frac{\nu^2}{2}\sum_{k=1}^{\infty}\left(k+\frac{1}{2}\right)t_k\frac{\partial F^{'}}{\partial
t_{k-1}}+\frac{z}{2}=0\ ,
  \end{equation}
leading to 
\begin{equation}
  \sum_{k=1}^{\infty}\left(k+\frac{1}{2}\right)t_kR_{k}-{z}\equiv {\cal R}=0\ ,
  \labell{eq:finally}
\end{equation}
where we have used the first integral of the  KdV flows~\reef{eq:kdvflows}, in the form:
\begin{equation}
\frac{\partial F^{'}}{\partial
t_{k-1}}=R_k \ .
  \labell{eq:kdvflowsagain}  
\end{equation}

The structure of these equations become even more natural when one
expresses it in conformal field theory language. There is a
$\IZ_2$--twisted boson with mode expansion:
\begin{equation}
  \labell{eq:modeexpansion}
  \partial\varphi(x)=\sum_{n}\alpha_{n+\frac{1}{2}}x^{-n-\frac{3}{2}}\ .
\end{equation}
The Virasoro generators are then modes of the stress tensor of this
boson, acting on a state for which the creation and annihilation
operators act by multiplication and derivation:
\begin{equation}
  \labell{eq:annihilate}
  \alpha_{-n-\frac{1}{2}}=\frac{1}{2\sqrt{2}\nu}\left(n+\frac{1}{2}\right)t_n\
  ,\qquad
  \alpha_{-n-\frac{1}{2}}={2\sqrt{2}\nu}\left(n+\frac{1}{2}\right)\frac{\partial}{\partial
  t_n}\ .
\end{equation}
Such states are coherent states:
\begin{equation}
  \labell{eq:coherent}
  <t|=<0|\exp\left(\sum_n t_n \alpha_{n+\frac{1}{2}}\right)\ ,
\end{equation}
and the partition function defines a special state $\Omega_{\bar t}$
in the Fock space of $\varphi(x)$:
\begin{equation}
  \labell{eq:Fock}
  \tau=<t|\Omega_{\bar t}>\ .
\end{equation}
It is clear now from equation~\reef{eq:loopmicro} and
equation~\reef{eq:modeexpansion} that our loop operator is the twisted
boson\cite{Dijkgraaf:1991rs}:
\begin{equation}
  \labell{eq:loopistwisted}
  w(x)\simeq\frac{1}{\nu}<0|\partial\varphi(x)|\Omega_{\bar t}>\ ,
\end{equation}
the square root branch cut we studied earlier realizing the
$\IZ_2$--twist.

The real bonus comes when we see how naturally the non--zero $\sigma$
and $\Gamma$ case fits into this formalism\cite{Johnson:1994vk}. The microscopic loop
equations are:
\begin{equation}
  \labell{eq:openloops}
  {\tilde L}_n\cdot \tau_{\{\sigma,\Gamma\}}=0\ ,
\end{equation}
where
\begin{equation}
  \labell{eq:openvirasoro}
  {\tilde
  L}_n=L_n-(1+n)\frac{\Gamma^2}{4}\sigma^n-\sigma^{n+1}\frac{\partial}{\partial\sigma}
  \ .
\end{equation}
In particular, if we use that the resolvent ${\hat R}_{\sigma,z}$ has an expansion in terms of the Gelfand--Dikii polynomials as
\begin{equation}
  \labell{eq:expansdresolvent}
  {\cal R}(\sigma,z)=\sum_{k=0}^\infty\frac{R_{k}}{\sigma^{k+\frac{1}{2}}}\ ,
\end{equation}
the first constraint is equivalent to equation~\reef{eq:nonpertbos},
(the rest following from the KdV flows) if the following relation is
true:
\begin{equation}
  \labell{eq:partial}
  \frac{\partial}{\partial\sigma}=-2\Gamma\nu\sum_{n=0}^{\infty}\sigma^{-k-\frac{3}{2}}\frac{\partial}{\partial
  t_k}\ ,
\end{equation}
which is equivalent to\cite{Itoh:1992vv,Johnson:1994vk}:
\begin{equation}
  \labell{eq:changetee}
  t_k={\tilde
  t}_k+\frac{2\Gamma\nu}{\left(k+\frac{1}{2}\right)}\, \,\sigma^{-k-\frac{1}{2}}\ .
\end{equation}
This relation is a preparation of a specific combination of closed
string operators, shifting the background. But from our studies in
earlier sections, we know that this background corresponds to
D--branes in an expansion in large positive $\mu$, and fluxes for type
0A in the large negative $\mu$ limit.

In fact, the Virasoro constraints~\reef{eq:openloops} with shifted
Virasoro operators~\reef{eq:openvirasoro} are precisely those of a
vertex operator of weight $\Gamma^2/4$,
$V_\Gamma=:e^{-\frac{\Gamma}{\sqrt{2}}\varphi(\sigma)}:$ and the
pleasingly simple relation can be derived\cite{Johnson:1994vk}:
\begin{equation}
  \labell{eq:tauopen}
  \tau_{\{\sigma,\Gamma\}}=<t|:e^{-\frac{\Gamma}{\sqrt{2}}\varphi(\sigma)}:|\Omega_{\bar t}>\ .
\end{equation}

Finally, note that the weight of the vertex operator is $\Gamma^2/4$.
The weight of the twisted boson itself is $1/16$. This means that when
$\Gamma=1/2$, the combined wavefunction given in
equation~\reef{eq:tauopen} comprising the partition function will have
vanishing weight, as can be seen by looking at the $n=0$ case of
equation~\reef{eq:openvirasoro} and~\reef{eq:constraints}. We do not
know the significance of this, but it is of note that for this case
there is a special exact non--perturbative solution of the string
equation~\reef{eq:nonpert} for any value of $k$, as displayed in
equation~\reef{eq:exact}.  It would be interesting to explore the role
of this solution further. Note that, for given $k$, there are $k+1$
solutions of this type, with $\Gamma=m+1/2$, $m=0,\ldots,k$, and 
the weight of the partition function then reduces to the interesting
quantity, $m(m+1)/4$.


\section*{Acknowledgments}
CVJ wishes to thank  James E. Carlisle for many questions and  discussions.

\providecommand{\href}[2]{#2}\begingroup\raggedright\endgroup

\end{document}